\documentclass[prd, twocolumn,amsmath]{revtex4}
\pdfoutput=1
\usepackage{hyperref}
\usepackage{amssymb}
\usepackage{graphicx}
\usepackage{color}

\newcommand{\avg}[1]{\left\langle {#1} \right\rangle}

\newcommand{\beq}{\begin{equation}}
\newcommand{\eeq}{\end{equation}}

\newcommand{\comment}[1]{}
\newcommand{\arxiv}[1]{arXiv: \href{http://arxiv.org/abs/#1}{#1}}
\newcommand{\doi}[1]{DOI: \href{http://dx.doi.org/#1}{#1}}

\begin{document}

	\title{Event Horizon Deformations in Extreme Mass-Ratio Black Hole Mergers}
	\author{Ryan Hamerly}
	\email{rhamerly@stanford.edu} % This address will be active throughout graduate school, while my Caltech address will expire in a year.
	\author{Yanbei Chen}
	\email{yanbei@tapir.caltech.edu}
	\affiliation{Theoretical Astrophysics 350-17, California Institute of Technology, Pasadena, California 91125}
	\date{December 7, 2011}
	
	\begin{abstract}
We study the geometry of the event horizon of a spacetime in which a small compact object plunges into a large Schwarzschild black hole. We first use the Regge-Wheeler and Zerilli formalisms to calculate the metric perturbations induced by this small compact object, then find the new event horizon by propagating null geodesics near the unperturbed horizon.  A caustic is shown to exist before the merger.  Focusing on the geometry near the caustic, we show that it is determined predominantly by large-$l$ perturbations, which in turn have simple asymptotic forms near the point at which the particle plunges into the horizon.  It is therefore possible to obtain an analytic characterization of the geometry that is independent of the details of the plunge.  We compute the invariant length of the caustic.  We further show that among the leading-order horizon area increase, half arises from generators that enter the horizon through the caustic, and the rest arises from area increase near the caustic, induced by the gravitational field of the compact object. 
	\end{abstract}

	\maketitle
% Some changes made in v.4 of the paper:
%
% 1. Changed some of the wording in the introduction.  Added figure of the horizon 'pant-leg' diagram.
% 2. 'space-time' -> 'spacetime', 'black-hole' -> 'black hole'
% 
% Things to look out for:
%
% 1. Consistency of wording: 'time-like' vs. 'timelike', etc.
% 2. Consistency of indexing: do we consistently use $H_0$ rather than $H_0^{lm}$, etc.?
% 3. Contravariant indices for forcing terms: $F_r$ should be $F^r$, etc.
% 4. Consistency in equations: (15-17) vs. (15)-(17)

%% Best to put this at the end of the paper.  Worry about later.
%	
%	\begin{acknowledgements}
%		I must of course thank my research advisor, Professor Yanbei Chen, who offered numerous insights through the course of this project.  Yanbei was able to think of new, interesting ways to look at problems, to which many
%		aspects of this thesis can be accredited.  Yanbei also had a firm command of the research literature and was able to link the work we did with both present and past developments in relativity theory.
%		
%		I should also thank Huan Yang, a graduate student in the TAPIR group, who did some work on this thesis and on many related topics.
%		
%		This thesis developed out of summer research project supported by the David and Judith Goodstein SURF Endowment.  I should give them thanks for supporting me during what was certainly my most enjoyable summer at Caltech.
%	\end{acknowledgements}
%

% Table of Contents needed?	
%	\tableofcontents
	
%	\mainmatter
	
	\section{Introduction}

The forthcoming prospects for gravitational-wave detection~\cite{GW} have motivated a great deal of study into black hole mergers, using both perturbative \cite{PN,BHP} and fully numerical methods~\cite{Pretorius}.  The most immediately relevant results are the gravitational waveforms emitted by such mergers, since accurate theoretical templates for these waveforms are crucial for their first detection \cite{HF}.   As a further step, however, studying features of spacetime geometry in the strong-field region, as well as their possible connection with out-going gravitational waves, will facilitate the use of gravitational waves as a tool for studying the non-linear and highly dynamical regime of geometrodynamics~\cite{Lovelace:2009,Keppel:2009,Nichols:2010,Rezzolla:2010}.  

In this work, we focus on the defining geometrical feature of a black hole spacetime, the event horizon --- a ``surface of last return'' which separates those points which can be connected to future infinity from those which cannot~\cite{HawkingPaper,PenroseBook}.  The horizon is a three-dimensional surface in a four-dimensional spacetime, it is a {\it null surface}, traced out by a two-parameter family of null geodesics.  At each point $\mathcal{P}$ on the event horizon, its tangent space $T_{\mathcal{P}}$ contains at least one {\it null generator}, which is tangent to a null geodesic whose entire future development lies on the horizon.  There also may exist points on the horizon, e.g., $\mathcal{Q}$ in the figure, at which two or more null vectors lie tangent to the horizon.  Although future developments of these null generators will stay on the horizon, their past developments do not belong to the event horizon.  Points like $\mathcal{Q}$ are called {\it caustic points}.

\begin{figure}[btp]
	\centering
		\includegraphics[width=0.75\columnwidth]{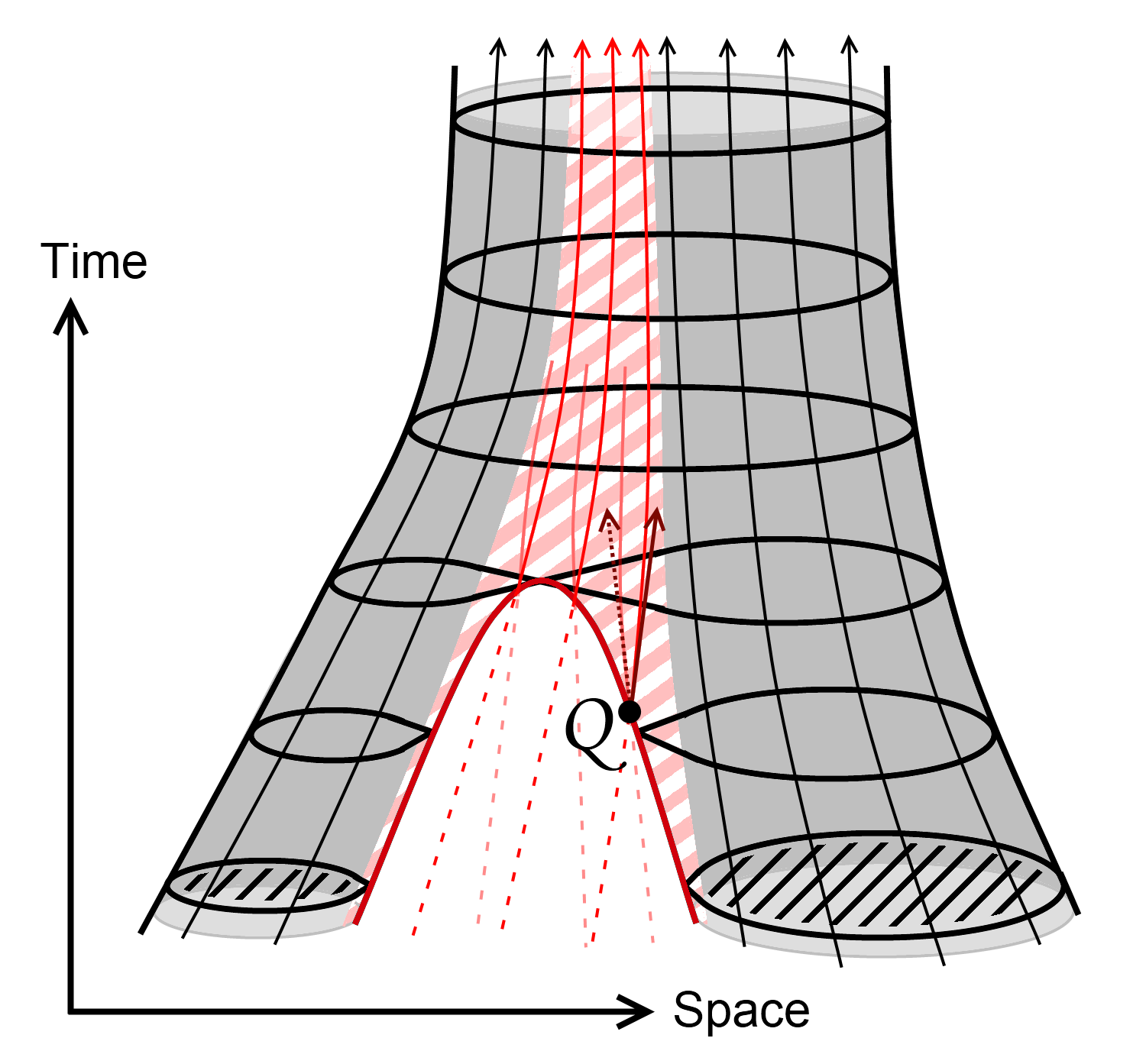}
	\caption{Spacetime diagram of the merger of two black holes.  The horizontal cross sections give the three-dimensional geometry of the event horizons as time progresses.  The null rays which trace out the horizon are
	given by the black and red directed lines.  The black lines originate from the horizon at past infinity, while the red lines \emph{enter} the horizon through the \emph{caustic}.  A caustic point $\mathcal{Q}$ is shown with its
	two null generators entering the horizon.}
	\label{fig:ResearchThesis-5}
\end{figure}

 By choosing a time slicing, one can take a three-dimensional cross-section of the spacetime to get the horizon's geometry at the present time -- a two-dimensional surface in a three-dimensional space.  To state an example, the event horizon of a static Schwarzschild black hole is a cylinder ($S^2 \times R$) in spacetime, or viewed in terms of its time slicings, it is a spherical surface ($S^2$) in space.  The horizon of a black-hole merger spacetime, on the other hand, has a ``pant-leg'' shape, which, in terms of its time slicings, looks like two roughly spherical objects.  These are the horizons of the merging black holes which, evolving as a function of time, merge to form a larger spherical object.  These time slicings of the horizon are often referred to as the event horizon itself.
A general operational way to obtain the event horizon, e.g., for the binary black-hole merger spacetime, is to first go to the final state of the spacetime, in which a final, nearly quiescent, black hole exists, with an easily identifiable, late portion of the event horizon.  Null rays on that horizon can be propagated backwards in time, and trace out the entire event horizon~\cite{Libson:1996,Masso:1999,CPS}. % I question putting these sentences in the same paragraph as the above, since it breaks the train of thought.  (Update: I moved this from the previous paragraph to this paragraph)

As it turns out, the existence of a caustic on the event horizon is quite general,  most notably in black hole merger spacetimes. Intuitively, a caustic develops {\it before the merger} because null generators, propagating backwards in time along the horizon of one of the merging black holes, are gravitationally lensed by the field of its companion, causing some of them to cross and necessarily leave the horizon, as is shown in Figure \ref{fig:ResearchThesis-5}. % Moved to its own paragraph.  Comments?
For black hole merger spacetimes, numerical simulations confirm this intuition~\cite{Winicour,HeadOn,CPS}, and moreover the mathematics of general relativity requires that a caustic form under generic merger conditions~\cite{SiinoPaper}. % Rephrase 'under generic merger conditions'?

Although perturbation theory has been applied to study the deformation of the event horizon due to tidal fields~\cite{Hartle:1974,Poisson:2010,Taylor:2008,Preston:2006,Preston:2006b}, as well as dissipation caused by this deformation~\cite{Comeau:2009,Poisson:2009,Poisson:1995,Fang:2005,Poisson:2004}, these previous works did not consider caustics on the horizon caused by plunging objects. In this paper, we study the caustic due to the plunge of a small point mass (with mass-energy $\mu$) into a big non-spinning black hole (with mass $M$), using perturbation theory.   

After working out the metric perturbations, we locate the new event horizon by propagating null geodesics around the future horizon of the unperturbed Schwarzschild spacetime.    Since the perturbation we apply is only valid to within a distance $\gg \mu$ from the small black hole (e.g., as measured in its local asymptotic rest frame), as we shall see later in the paper, our study will have to exclude a region in the event horizon that has area of approximately $O(\mu^2)$ towards the final future horizon.   Fortunately, most of the change in geometry is caused by rays that travel at a distance $\sim \sqrt{ \mu M } \gg \mu$ from the small hole,  where gravity is still weak.  More specifically, since the final hole will have a mass of $M[1+\mu/M+O(\mu^2/M^2)]$ and angular momentum of $O(\mu M)$,  its area will be $16\pi^2M^2[1+2\mu/M+O(\mu^2/M^2)]$ --- linear black-hole perturbation will account for the leading order area increase due to the plunge of the black hole, in particular the contribution due to those rays that enter through the caustic.  Moreover,  as it turns out, rays significantly influenced by the small black hole all tend to go close enough to the small black hole, such that the small black hole's influence can be approximated as ``instantaneous''.   This allows us to develop an impulse approximation (also called the Born Approximation in scattering theory~\cite{BornApproximation}) that leads to an analytic description of the caustic and geometry around it.

This paper will be organized as follows:  In Sec.~\ref{sec:metricandhorizon}, we first briefly review the Regge-Wheeler and Zerilli formalisms %should this be plural?
 for black-hole metric perturbations, and then apply these results to the propagation of null geodesics near the future horizon, writing down their evolution equations, which are driven by ``forcing terms'', which can in turn be written in terms of Zerilli and Regge-Wheeler functions.  In Sec.~\ref{sec:delta-function}, we develop an {\it impulse approximation} for null-ray propagation and show that the deflection of null rays depends only on the time integral of the forcing terms, instead of their detailed dependence on time.  In Sec.~\ref{sec:caustics}, we apply the impulse approximation to the propagation of null rays close to the unperturbed horizon, obtaining the caustic structure of the new horizon.  In Sec.~\ref{sec:area}, we calculate the event-horizon area change due to new rays that enter the horizon through the caustic and due to increase in area induced by the small object.  
 %In Sec.~\ref{sec:impulseshape}, we compute the shape of the impulse using perturbation theory and demonstrate, as we would expect, that it resembles the transverse acceleration of a light ray scattered in the gravitational field of a mass $\mu$.  
In Sec.~\ref{sec:strings}, we relax the point-particle assumption and treat the case of one-dimensional ``strings'' falling into the black hole.  This results in significant changes to the caustic geometry and the distribution for area increase.  In Sec.~\ref{sec:conclusions}, we summarize our main conclusions. 
% And don't forget to add a description of the l = 1 section if we write a section dedicated to l = 1 terms (i.e. frame-dragging of caustic, etc.)

	\section{Metric perturbation and deformation of the event horizon} 
	\label{sec:metricandhorizon}
	
	\subsection{Metric Perturbations: the Regge-Wheeler Gauge}
	\label{sec:metric}
	
		Suppose $\mu \ll M$, and using the geometrical units of $G=c=M=1$, we consider a background Schwarzschild spacetime with the black hole mass set to unity:
		\beq
			ds^2 = -(1-2/r)dt^2 + (1-2/r)^{-1}dr^2 + r^2 d\Omega^2
		\eeq
The small black hole's world line in this background spacetime, up to leading order in $\mu$, is a timelike geodesic.  We consider a first-order perturbation (at $O(\mu)$ order) induced on this background spacetime by the small black hole. Dependence of the metric perturbations on angular coordinates $\theta$ and $\phi$  can be decomposed into scalar, vector and tensor harmonics, and classified according to parity.  As shown by Regge and Wheeler~\cite{ReggeWheeler}, a choice of gauge (the Regge-Wheeler gauge) allows us to eliminate all but 6 fields, and write 
		\begin{eqnarray}
			ds_p^2 & = & \left[(1-2/r)H_0^{lm} dt^2 + 2H_1^{lm} dtdr + \right.\nonumber \\
			 	& & \left. H_2^{lm}(1-2/r)^{-1} dr^2 + r^2 K^{lm} d\Omega^2\right]Y^{lm} \nonumber \\
				& & + 2h_0^{lm} \left[\sin\theta Y^{lm}_{,\theta}dtd\phi - \csc\theta Y^{lm}_{,\phi}dtd\theta\right] \nonumber \\
				& & + 2h_1^{lm} \left[\sin\theta Y^{lm}_{,\theta}drd\phi - \csc\theta Y^{lm}_{,\phi}drd\theta\right] \label{eq:metricpert}	
		\end{eqnarray}
		Here $(H_0, H_1,H_2,K)$ are ``even-parity'' perturbations with a parity of $(-1)^l$, and $(h_0,h_1)$ are ``odd-parity'' perturbations, with parity of $(-1)^{l+1}$.   Henceforth in the paper, we drop the $(lm)$ dependence of
		all metric-perturbation fields.

		Regge and Wheeler~\cite{ReggeWheeler} deduced 10 linearized Einstein Equations for these 6 fields. Among these, 7 are even-parity and 3 are odd-parity.  Zerilli~\cite{ZerilliPaper} showed that the monopole and dipole
		perturbations can be found exactly by integrating the equations of motion.  For multipoles with $l \geq 2$, however, it is not clear from the Einstein equations whether one can solve for the metric perturbations
		systematically, e.g., as an initial-boundary-value problem.  However, for odd and even-parity perturbations respectively, Regge-Wheeler~\cite{ReggeWheeler} and Zerilli~\cite{ZerilliLetter} were able to construct functions $Q$
		(often referred to as the Regge-Wheeler function) and $Z$ (often referred to as the Zerilli function) that satisfy wave equations in vacuum, which can be solved using standard approaches --- and all metric perturbation fields
		can then be expressed in terms of $Z$ and $Q$.  Zerilli~\cite{ZerilliPaper} further worked out the source terms that appear on the right-hand side of the wave equations, when a point particle falls along a geodesic
		--- as well as modifications that must be made in the relations between $Z$ and $Q$ and metric perturbation fields.  By these procedures, we can find the metric perturbations for all $l$ values.
		
		\subsection{Perturbations with $l \geq 2$}
			We briefly review the procedures used to obtain these perturbations in Appendices~\ref{sec:even} (for even parity) and \ref{sec:odd} (for odd parity).   Here we simply list the conversion equations:
			\begin{eqnarray}
				\label{eq:H2toH0} H_0 & = & H_2 + S_{H_0} \\
				\label{eq:ZtoH2} H_2 & = & -\frac{r^3\lambda^2(\lambda+1) + 3r^2\lambda^2 + 9r\lambda + 9}{r^2(\lambda r+3)^2} Z \nonumber \\
					& & + \frac{r^2\lambda - 3r\lambda - 3}{(r-2)(\lambda r+3)}\frac{\partial Z}{\partial r_*} + \frac{r^2}{r-2} \frac{\partial^2 Z}{\partial r_{*}^2} + S_{H_2} \\
				\label{eq:ZtoH1} H_1 & = & \frac{\lambda r^2 - 3\lambda r - 3}{(r-2)(\lambda r+3)} \frac{\partial Z}{\partial t} + \frac{r^2}{r-2} \frac{\partial^2 Z}{\partial r_* \partial t} + S_{H_1} \\
				\label{eq:ZtoK} K & = & \frac{\lambda(\lambda+1)r^2 + 3\lambda r + 6}{r^2(\lambda r+3)} Z + \frac{dZ}{dr_*} + S_{K} \\
				\label{eq:Ztoh0} h_0 & = & \frac{r-2}{r} \int{Q dt} + r\int{\frac{\partial Q}{\partial r_*}dt} + S_{h_0} \\
				\label{eq:Ztoh1} h_1 & = & \frac{r^2}{r-2} Q
			\end{eqnarray}
			where
			\beq
				\lambda \equiv \frac{1}{2}(l-1)(l+2)
			\eeq
			and the wave equations,
			\begin{eqnarray}
				\label{eq:zerilli} \frac{\partial^2 Z}{\partial r_*^2} - \frac{\partial^2 Z}{\partial t^2} - V_l^Z(r) Z & = & S^Z_{lm} \\
				\label{eq:zerilliodd} \frac{\partial^2 Q}{\partial r_*^2} - \frac{\partial^2 Q}{\partial t^2} - V_l^{Q}(r) Q & = & S_{lm}^{Q}
			\end{eqnarray}
			The quantities $S_{H_0}$ through $S_{h_0}$ are  placeholders for the source terms in Equations (\ref{eq:b62}), (\ref{eq:k}-\ref{eq:h2}), and (\ref{eq:h0odd}-\ref{eq:h1odd}).  The source terms $S_{lm}^Z$ and $S_{lm}^{Q}$
			take rather cumbersome forms and have been consigned to Equations (\ref{eq:zersrc}) and (\ref{eq:zersrcodd}) of the Appendix.  The potential terms in the Zerilli and Regge-Wheeler wave equations are
			\begin{eqnarray}
				V_l^Z(r)\!\! & = &\!\! 2 \left(1-\frac{2}{r}\right) \frac{\lambda^2(\lambda+1)r^3 + 3\lambda^2 r^2 + 9\lambda r + 9}{r^3(\lambda r+3)^2} \quad\\
				V_l^Q(r)\!\! & = & \!\! \left(1-\frac{2}{r}\right)\left[\frac{2(\lambda+1)}{r^2} - \frac{6}{r^3}\right] 
			\end{eqnarray}
			We have also defined
			\begin{equation}
				r_*  =   r + 2 \ln(r/2 - 1)\,,
			\end{equation}
			which is often referred to as the tortoise coordinate.
	
			Since we are interested in the deformation of the {\it future event horizon}, we specialize the relations (\ref{eq:H2toH0}-\ref{eq:Ztoh1}) to the event horizon, where $(r-2)\rightarrow 0$ and $Z$ and $Q$ are functions of 
			$v \equiv t+r_*$ alone: 
			\begin{eqnarray}
				\label{eq:keq} K & = & \frac{8\pi \mu Y_{lm}^{*(0)}}{2\lambda + 3} \Theta(v) + \left(\frac{\lambda+1}{2} + \frac{d}{dv}\right)Z \\
				H_0 & = & H_1 = H_2 \nonumber \\
					\label{eq:h1eq} & = & \frac{4}{r-2}\left[\frac{8\pi \mu Y_{lm}^{*(0)}}{2\lambda + 3} \delta(v) + \left(\frac{d}{dv} - \frac{1}{4}\right)\frac{dZ}{dv}\right] \\
				\label{eq:h0oddeq} h_0 & = & 2 Q \\
				\label{eq:h1oddeq} h_1 & = & \frac{4}{r-2} Q
			\end{eqnarray}
			where $\Theta(v)$ is the Heaviside step function, $\mu = m_0 E$ is the mass-energy of the small hole in the Schwarzschild background ($m_0$ is rest mass and $E$ the specific energy associated with its geodesic world line),
			and	$Y_{lm}^{*(0)} \equiv Y_{lm}^*(\theta_0,\phi_0)$ is the complex conjugate of the spherical harmonic taken at angular coordinates $(\theta_0, \phi_0)$ where the particle meets the horizon.
			
			As we can see from Eqs.\ (\ref{eq:keq}-\ref{eq:h1oddeq}), the metric perturbations are given by ingoing waves (i.e. functions of $v$) multiplied by various powers of $(r-2)$.  We may write this dependence more explicitly 
			as:
			\begin{eqnarray}
				K = \mathcal{K}(v) & & H_0 = \frac{\mathcal{H}_0(v)}{r-2} \nonumber \\
				H_1 = \frac{\mathcal{H}_1(v)}{r-2} & & H_2 = \frac{\mathcal{H}_2(v)}{r-2} \label{eq:pert-asy} \\
				h_0 = \mathfrak{h}_0(v) & & h_1 = \frac{\mathfrak{h}_1(v)}{r-2} \nonumber
			\end{eqnarray}
			where $\mathcal{K}$, $\mathcal{H}_0$, $\mathcal{H}_1$, $\mathcal{H}_2$, $\mathfrak{h}_0$, and $\mathfrak{h}_1$ are defined to make % (\ref{eq:keq2}-\ref{eq:h1oddeq2})
			(\ref{eq:pert-asy}) consistent with (\ref{eq:keq}-\ref{eq:h1oddeq}).
		
		\subsection{Low Multipole ($l<2$) Perturbations}

For low values of $l$, not all Regge-Wheeler fields are involved in parametrizing the full metric perturbation.  The linearized Einstein Equation, consequently, will be dramatically simplified.  For example, when $l=0$, vector and tensor harmonics all vanish, while for $l=1$, only the tensor harmonics vanish. A direct mathematical consequence is that a wave equation cannot be constructed for perturbations with $l<2$ --- yet equations are simpler so that they can be solved directly.    In Appendix G of Ref.~\cite{ZerilliPaper}, Zerilli provides solutions to all these cases for a point particle perturbing a Schwarzschild space-time.  We do not  repeat his derivation, but merely state the results.

		\subsubsection{Monopole $(l=0)$ Term}
% Zerilli has a lot of errors in his paper, including a misplaced factor of i (which makes his metric perturbation imaginary) and a misplaced factor of -1.  I have rederived the correct formulas below.
% Should I repeat Zerilli's derivation somewhere?  
		The monopole perturbation ($l = 0$) is associated with the mass of the black hole.  At a distance $r$, it is related to the amount of mass enclosed within the coordinate sphere with radius $r$.  Since vector and tensor harmonics do not exist for $l=0$, the only perturbation fields are $H_0$, $H_1$ and $H_2$. As Zerilli has shown in Appendix G of Ref.~\cite{ZerilliPaper},  after a gauge transformation, the only surviving even-parity terms are $H_0$ and $H_2$.  For a plunging particle of mass-energy $\mu$, we have:
			\beq
				H_0 = H_2 = \frac{2\sqrt{4\pi} \mu}{r-2} \Theta(v)  % This differs from Zerilli's notation by a minus sign, and Zerilli's notation is clearly wrong here (maybe they used the wrong metric signature?)
			\eeq
			or equivalently,
			\beq
				\label{eq:monopole-h} \mathcal{H}_0 = \mathcal{H}_2 = 2\sqrt{4\pi} \mu \Theta(v)
			\eeq
			where $\Theta(v)$ is the Heaviside step function.
			
			\subsubsection{Odd-Parity Dipole $(l=1)$  Term}
			
			The odd-parity dipole perturbation represents the spin imparted by the small black hole's orbital angular momentum.  This is slightly less trivial than the $l = 0$ case, but can be simplified by a gauge choice which makes
			$h_1$ vanish and $h_0$ approach a constant in time both before and after the plunge.  A gauge can be chosen in such a way that metric perturbation before the plunge vanishes, and after the plunge acquires a value that
			depends on the orbital angular momentum $\vec{L}$ of the plunging particle:
			
			\begin{eqnarray}
				\delta g_{tr} & = & 0 \\
				\delta g_{t\theta} & = & \frac{2}{r} \left[L_x \sin\phi - L_y \cos\phi\right] \\
				\delta g_{t\phi} & = & \frac{2}{r} \left[-L_z \sin^2\theta + (L_x \cos\theta \right.\nonumber \\
					& & \left. + L_y \sin\theta) \sin\theta\cos\theta\right]
			\end{eqnarray}
			
			This corresponds to a metric perturbation of:
			\begin{eqnarray}
				h_0^{1m} & = & \mathfrak{h}_0^{1m} = \sqrt{4\pi/3} \frac{2 m_0 L^m}{r} \Theta(t-T(r)) \nonumber \\
					& \stackrel{r \rightarrow 2}{\rightarrow} & \sqrt{4\pi/3}\,m_0 L^m \Theta(t-T(r))
			\end{eqnarray}
			where $L^m$ is the spherical-harmonic representation % wording?
			of the plunging orbital angular momentum $\vec{L}$ -- i.e. $L^0 = L^z$ and $L^{\pm1} = 2^{-1/2}(L_x \pm i L_y)$

%\textcolor{red}{What is this L here? It must be the orbital angular momentum of the particle. Therefore, we only need to consider $L_z$.   In fact, our orbit only has $L_z$. } Orbit can have more general L's if we don't choose our axes carefully.

			\subsubsection{Even-Parity Dipole Term}

			In the absence of a source, the even-parity dipole term is a gauge that can be eliminated by transforming to the center-of-mass frame, in which the center of mass lies at rest at the origin of the coordinate system. When a source is present, however, the term cannot be eliminated, since it is not 
			possible to gauge away a source term (although a gauge tranformation \emph{can} concentrate the even-parity term along the path of the particle).  Choosing to work in a gauge where $K = 0$, the Einstein Equations can be integrated exactly to give:
			\begin{eqnarray}
				H_0^{1m} & = & \frac{f_{1m}(t) + r^3 f_{1m}''(t)}{(r-2)^2}\Theta(r-R(t)) \\
				H_1^{1m} & = & -\frac{r f_{1m}'(t)}{(r-2)^2}\Theta(r-R(t)) \\
				H_2^{1m} & = & \frac{f_{1m}(t)}{(r-2)^2}\Theta(r-R(t))
			\end{eqnarray}
			where
			\beq
				f_{1m}(t) = 8\pi\mu (R(t)-2) Y_{1m}^*
			\eeq
			Near the horizon, we have $(R(t) - 2) \sim e^{-t/2}$, so $f_{1m}' = -\frac{1}{2} f_{1m}$ and $f_{1m}'' = \frac{1}{4} f_{1m}$.  In Section (\ref{sec:horizon}), where we trace out the structure of the perturbed horizon, we
			will make use of the values of the metric coefficients, as functions of $v$, along lines of constant $u = t - r_*$ (where $(r - 2) \sim e^{v/4}$ and $(R(v)-2) \sim e^{-v/4}$).  Doing so here, and setting $v = 0$ to
			correspond to the point at which $R(v)$ crosses the line of constant $u$ (so $r-2 = R(v)-2$ at $v = 0$), we find:
			\begin{eqnarray}
				\mathcal{H}_0^{1m} & = & \mathcal{H}_1^{1m} = \mathcal{H}_2^{1m} \nonumber \\
					& = & 8\pi\mu \frac{R(v)-2}{r-2} \Theta(v) Y_{1m}^{*} \nonumber \\
					& = & 8\pi\mu e^{-v/2}\Theta(v) Y_{1m}^{*}  \label{eq:dipole-h}
			\end{eqnarray}
			Of importance, we note that all metric coefficients vanish at past and future infinity, indicating that, in these limits, the coordinate frame is centered around the large black hole.  Since it is the large black hole that we are interested in, this is the proper coordinate frame to use.

	\subsection{The Deformed Event Horizon} \label{sec:horizon}
		
		In order to analyze the horizon deformation caused by metric perturbation fields (\ref{eq:keq}-\ref{eq:h1oddeq}), we need to study the propagation of light rays near the horizon.  We will do so in the light-cone Kruskal-Szekres coordinates, which offer the distinct advantage of having non-singular light cones around the horizon. 
		The Kruskal coordinates 
		$(V, U, \theta, \phi)$ are related to their Schwarzschild counterparts by:
		\begin{eqnarray}
			X^0 \equiv V & = & e^{v/4} \\
			X^1 \equiv U & = & -e^{-u/4} \\
			X^2 & = & \theta \\
			X^3 & = & \phi
		\end{eqnarray}
		where
		\begin{eqnarray}
			v & = & t + r_* \\
			u & = & t - r_* 
		\end{eqnarray}
		We model the unperturbed horizon as a set of null generators, parameterized  by $V$.  To distinguish these horizon generators from the rest of the null rays in the system, we impose the {\it final condition}
		$U \rightarrow 0$ as $V \rightarrow \infty$.  The generators of the unperturbed horizon are then given by:
		\begin{eqnarray}
			V(V) & \equiv & X^{0}(V) = V \\
			U(V) & \equiv & X^{1}(V) = 0 \\
			\theta(V) & \equiv & X^{2}(V) = \mbox{const} \\
			\phi(V) & \equiv & X^{3}(V) = \mbox{const}
		\end{eqnarray}
		These generators satisfy the geodesic equation, modified to account for the non-affine parametrization:
		\beq
			\frac{d^2X^\mu}{dV^2} = -\bar{\Gamma}^\mu_{\nu\rho} \frac{dX^\nu}{dV} \frac{dX^\rho}{dV} + g \frac{dX^\mu}{dV}
		\eeq
		Here, $\bar{\Gamma}^\mu_{\nu\rho}$ refers to the Kruskal Christoffel symbol. 
		The non-perturbed horizon at $U=0$ indicates 
		\begin{eqnarray}
			\label{eq:geq} g & = & \bar{\Gamma}^0_{00}  = 0\,,\quad 
			\bar{\Gamma}^i_{00}=0\,. 
		\end{eqnarray}
		Note that the unperturbed event-horizon is affine parametrized.
		
		On a perturbed metric, the rays themselves will be perturbed; supposing we still parametrize the horizon by $V$, then we need to modify
		$X^j \rightarrow X^j + \delta X^j$ ($j=1,2,3$) and $g
		\rightarrow g + \delta g$.  To first order in the metric perturbation, we obtain the following equations of motion:
		\begin{eqnarray}
			\label{eq:dgeq} \delta g & = & -2\bar{\Gamma}^0_{0i}\frac{d(\delta X^i)}{dV} - \delta\bar{\Gamma}^0_{00} - \bar{\Gamma}^0_{00,i}\delta X^i \\
			 \frac{d^2(\delta X^i)}{dV^2} & = & -2\bar{\Gamma}^i_{0j} \frac{d(\delta X^j)}{dV} - \delta \bar{\Gamma}^i_{00} - \bar{\Gamma}^i_{00,j}\delta X^j  + g \frac{d(\delta X^i)}{dV} \nonumber \\
			 \label{eq:dxeq}
		\end{eqnarray}
		Note in this case that since $\delta g \neq 0$, the perturbed event horizon is no longer affine-parameterized by $V$. 
		A careful derivation shows that all of the Christoffel symbols in (\ref{eq:geq}-\ref{eq:dxeq}) are finite in the limit $r \rightarrow 2$.  This is necessary in order for the perturbation theory to be well-posed.  Moreover,
		most of these Christoffel symbols (but not their perturbations) vanish outright, yielding the following equations of motion for the angular coordinates $\theta(V)$ and $\phi(V)$:
%		\begin{eqnarray}
%			\label{eq:ddu2} \frac{d^2(\delta\theta)}{dV^2} & = & -\delta\bar{\Gamma}^2_{00} \\
%			\label{eq:ddu3} \frac{d^2(\delta\phi)}{dV^2} & = & -\delta\bar{\Gamma}^3_{00}
%		\end{eqnarray}		\begin{eqnarray}
		\beq
			\label{eq:ddu2} \frac{d^2(\delta\theta)}{dV^2} = -\delta\bar{\Gamma}^2_{00},
			\ \ \ \ \frac{d^2(\delta\phi)}{dV^2} = -\delta\bar{\Gamma}^3_{00}
		\eeq

		In principle, we could have also derived an equation for $\delta U(V)$.  However, as we will see in Section \ref{sec:caustics}, in the region of interest, i.e.\ near the caustic, the angular perturbations will scale as $O(\mu^{1/2})$, while the radial perturbations scale  with the higher power $O(\mu)$.  It will not suffice to compute the geodesic equation to first order in the perturbation; in a rigorous treatment, higher-order perturbation terms would be needed here.
		
		Fortunately, there is another way to derive the radial equation that does not involve a cumbersome higher-order perturbation expansion.  The horizon generators are null rays, and always will be no matter what spacetime they propagate through.  Given $\theta(V)$ and $\phi(V)$ from (\ref{eq:ddu2}), we can derive an equation for $\delta U(V)$ by setting $g_{\mu\nu}u^\mu u^\nu = 0$:
		\beq
			\label{eq:ddu1new}\frac{1}{e} \frac{d(\delta U)}{dV} = \frac{1}{16}\delta\bar{g}_{00} + \frac{1}{4}\left[\left(\frac{d(\delta\theta)}{dV}\right)^2 + \sin^2\theta \left(\frac{d(\delta\phi)}{dV}\right)^2\right]
		\eeq
		The equations in their present form are solvable but cumbersome.  One can rewrite them in a more intuitive form by transforming into the ingoing Eddington-Finkelstein form, substituting $U$ for $r$ and writing $V$ in terms of $v$.  Upon simplification, equations
		(\ref{eq:ddu2}-\ref{eq:ddu1new}) transform into:
		\begin{eqnarray}
			\label{eq:geo-r} \left(\frac{d}{dv} - \frac{1}{4}\right) \delta r & = & F^r - 2\left[\dot{\theta}^2 + \dot{\phi}^2 \sin^2\theta \right] \\
			\label{eq:geo-th}\left(\frac{d}{dv} - \frac{1}{4}\right) \frac{d}{dv} \theta & = & F^\theta \\
			\label{eq:geo-ph}\left(\frac{d}{dv} - \frac{1}{4}\right) \frac{d}{dv} \phi & = & F^\phi
		\end{eqnarray}
		where $\dot{A}$ means $dA/dv$ for any $A$, and $F^r$, $F^\theta$, and $F^\phi$ are ``forcing terms'' that arise from the small black hole's perturbing field.
		
	\subsection{The Forcing Terms}
		The ``forcing terms'' $F^r$, $F^\theta$, and $F^\phi$ tell the rays on the horizon how far to bend in the small hole's gravitational field.  In terms of the Kruskal-frame metric perturbations and Christoffel symbols, they are given by:
		\begin{eqnarray}
			F^r & = & -\frac{V^2}{32} \delta\bar{g}_{00} \\
			F^\theta & = & -\frac{V^2}{16} \delta\bar{\Gamma}^2_{00} \\
			F^\phi & = & -\frac{V^2}{16} \delta\bar{\Gamma}^3_{00}
		\end{eqnarray}
		While the Christoffel symbol is not a tensor, its perturbation $\delta\Gamma^\mu_{\nu\rho}$ is.  The Kruskal-coordinate components of this tensor are related to its Schwarzschild components by the coordinate transformation:
		\beq
			\delta\bar{\Gamma}^\mu_{\nu\rho} = \frac{\partial X^\mu}{\partial x^\alpha} \delta\Gamma^\alpha_{\beta\gamma} \frac{\partial x^\beta}{\partial X^\nu} \frac{\partial x^\gamma}{\partial X^\rho}
		\eeq
		The Schwarzschild Christoffel symbols
		\beq
			\delta\Gamma^\mu_{\nu\rho} = \frac{1}{2} g^{\mu\sigma}(\delta g_{\sigma\nu;\rho} + \delta g_{\sigma\rho;\nu} - \delta g_{\nu\rho;\sigma})
		\eeq
		are calculated using the metric in
		(\ref{eq:metricpert}) and substituting the horizon metric perturbations (\ref{eq:keq}-\ref{eq:h1oddeq}).  This leads to the following forcing terms:
		\begin{eqnarray}
			\label{eq:fr} F^r & = & -\frac{1}{4}f^{\rm (e)}_{lm} Y^{lm} \\
			\label{eq:fth} F^\theta & = & \frac{1}{16} \left[f^{\rm (e)}_{lm}  Y^{lm}_{,\theta}  +  f^{\rm (o)}_{lm} \frac{Y^{lm}_{,\phi}}{\sin\theta}\right] \\
			\label{eq:fph} F^\phi & = & \frac{1}{16} \left[f^{\rm (e)}_{lm} \frac{Y^{lm}_{,\phi}}{\sin^2\theta}-f^{\rm (o)}_{lm}\frac{Y^{lm}_{,\theta}}{\sin\theta}\right]
		\end{eqnarray}
		Here we have defined 
		\begin{eqnarray}
			\label{eq:flme0} f_{lm}^{\rm (e)} & = & \frac{1}{4} \left[\mathcal{H}_0 + 2\mathcal{H}_1 + \mathcal{H}_2\right] \\
			\label{eq:flmo0} f_{lm}^{\rm (o)} & = & \left(\frac{d}{dv} - \frac{1}{4}\right) \left[2\mathfrak{h}_0 + \mathfrak{h}_1\right]
		\end{eqnarray}
		In terms of the Zerilli and Regge-Wheeler functions, we may write this as:
		\begin{eqnarray}
			\label{eq:flme} f_{lm}^{\rm (e)} & = & \frac{32\pi \mu Y_{lm}^{*(0)}}{2\lambda + 3} \delta(v) + 4\left(\frac{d}{dv} - \frac{1}{4}\right)\frac{dZ}{dv} \\
			\label{eq:flmo} f_{lm}^{\rm (o)} & = & 8\left(\frac{d}{dv} - \frac{1}{4}\right)Q
		\end{eqnarray}

	\section{Impulse Approximation} \label{sec:delta-function}
		Up to this stage, a straightforward calculation using the techniques developed above can already compute the metric perturbations, the shape of the horizon, and hence the structure of the caustic.  
		This involves numerically solving for the Zerilli and Regge-Wheeler functions for particular geodesics, and inserting them into Eqs.\ (\ref{eq:geo-r}-\ref{eq:geo-ph}) and (\ref{eq:fr}-\ref{eq:flmo}).  In particular, the
		wave equations can be solved either in the time domain, or in the frequency domain.  For example, for $Z$, we have 
		\begin{equation}
			\left[\omega^2 +\frac{\partial^2}{\partial r_*^2}  - V_l^Z \right] \tilde Z(\omega,r) = \tilde S^Z_{lm}\,, 
		\end{equation}
		with out-going boundary condition
		\begin{equation}
			\tilde Z(\omega,r_*) \sim e^{\pm i\omega r_*}\,, \quad r_*\rightarrow \pm\infty\,.
		\end{equation}
		However, when we look on the short lengthscales that characterize the caustic, or conversely if we look on the long timescales that characterize the initial and final states of the big black hole, the results can be greatly
		simplified by approximating each forcing term as an instantaneous impulse.

		\subsection{The Approximation}
			Consider a generator on the large black hole's horizon which is deflected by the small hole's gravitational field.  Let $\theta$ denote the impact parameter of this scattering problem.  Most of the deflection will occur
			within a time interval
			\beq
				\tau \sim \theta
			\eeq
			Define a ``boundary'' $\theta_b$ such that $\sqrt{\mu} \ll \theta_b \ll 1$.  All rays with $\theta < \theta_b$ are considered to be in the {\it vicinity of the caustic}, while rays with $\theta > \theta_b$ are considered to
			be in the {\it bulk} of the horizon, $\theta_b$ functioning as the boundary between these two regions.  We consider the impulse approximation for both cases separately:
			
			\subsubsection{Vicinity of the Caustic}
				As we will show in the next section, the geometry of the caustic depends only on those rays with $\theta \lesssim \sqrt{\mu} \ll \theta_b$ -- that is, rays in the {\it vicinity of the caustic}.  The deflection timescale
				for these rays is thus $\ll 1$, and the deflection can be approximated as an instantaneous impulse.  This is the intuitive essence of the impulse approximation; we discuss it in more rigorous detail below.
			
				In Appendices \ref{sec:zbehavior} and \ref{sec:qbehavior}, we show that $Z$ and $Q$ satisfy scaling relations for large $l$, namely:
				\beq
					Z_{lm} \sim l^{-3} Y_{lm}^{*(0)} f(v/l^{-1}),
					\ \ \ Q_{lm} \sim l^{-3} Y_{lm,\theta}^{*(0)} g(v/l^{-1})
				\eeq
			
\begin{figure}[tbp]
					\centering
						\includegraphics[width=1.00\columnwidth]{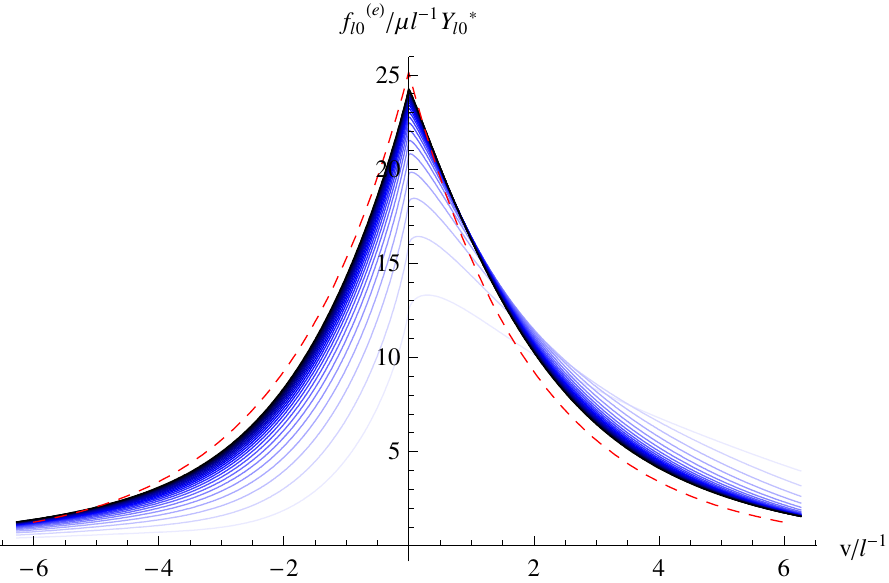}
					\caption{Plot of $f_{lm}^{(e)}/\mu l^{-1}$ versus $v/l^{-1}$ for a radial plunge, values $2 < l < 24$, $m = 0$ shown.  Larger values of $l$ are denoted by darker lines.  The red dashed line is the empirical limit
						curve $8\pi e^{-x/2}$; the significance of this curve is touched on in Section \ref{sec:impulseshape}.}
					\label{fig:ResearchPaper2-Fig1}
				\end{figure}

				The forcing terms (\ref{eq:flme}-\ref{eq:flmo}) depend on the Zerilli and Regge-Wheeler functions, and therefore satisfy similar scaling relations.  As an illustration, the scaling relation for the even term
				$f_{lm}^{(e)}$ is plotted in Figure \ref{fig:ResearchPaper2-Fig1}.  As $l$ increases, the forcing term increasingly resembles an impulse.  The impulse approximation thus replaces $f_{lm}^{(e)}$ and $f_{lm}^{(o)}$ with
				delta functions in time:
				\begin{eqnarray}
					f_{lm}^{\rm (e)}(v) & \rightarrow & \left[\int_{-\infty}^{\infty}f_{lm}^{\rm (e)}(v') dv'\right] \delta (v) \equiv \bar f_{lm}^{\rm (e)} \delta (v) \\
					f_{lm}^{\rm (o)}(v) & \rightarrow & \left[\int_{-\infty}^{\infty}f_{lm}^{\rm (o)}(v') dv'\right] \delta (v) \equiv \bar f_{lm}^{\rm (o)} \delta (v)
				\end{eqnarray}
			
			\subsubsection{The Bulk} \label{sec:impulse-bulk}
				We will prove in Section \ref{sec:caustics} that $\delta\theta \sim \mu/\theta$; let us for now assume {\it a priori} that this holds.  The deflection of rays in the bulk is then infinitesimal -- $\delta\theta \ll \theta$.
				Taking this to be the case, the forcing terms in (\ref{eq:geo-r}-\ref{eq:geo-ph}) lose their dependence on the deflection and become functions of time alone.  We can then integrate (\ref{eq:geo-r}-\ref{eq:geo-ph}) to
				give:
				\begin{eqnarray}
					\left.-\frac{1}{4}\delta r\right|^{+\infty}_{-\infty} & = & \left.F^r\right|^{+\infty}_{-\infty} \\
					\left.\frac{d\theta}{dv} - \frac{1}{4}\theta\right|^{+\infty}_{-\infty} & = & \int_{-\infty}^\infty{F^\theta(v)dv} \\
					\left.\frac{d\phi}{dv} - \frac{1}{4}\phi\right|^{+\infty}_{-\infty} & = & \int_{-\infty}^\infty{F^\phi(v)dv}
				\end{eqnarray}
				On long timescales $|v| \gg 1$, the angular perturbations depend only on the {\it time integral} of the forcing term; because of this, we can replace the forcing term with an instantaneous impulse.  The radial term,
				conversely, depnds on the value of $F^r$ at $v = \pm\infty$; as we will show, this gives rise to an increase in the black hole's radius by an amount $\delta r = 2\mu$.

		\subsection{The Impulse}
			\subsubsection{Even Parity, $l \geq 2$}
				Integrating $f_{lm}^{\rm (e)}$ over time with the help of (\ref{eq:flme}), we find:
				\beq
					\label{eq:imp-e1} \bar f_{lm}^{\rm (e)} = \frac{32\pi\mu{Y_{lm}^*}^{(0)}}{2\lambda+3} + \left.4\left(\frac{d}{dv} - \frac{1}{4}\right)Z\right|^{+\infty}_{-\infty}
				\eeq
				The Zerilli function approaches a constant as $v \rightarrow \pm\infty$, so the $dZ/dv$ term drops out, but the jump in $Z$ contributes to the final result.  In Appendix \ref{sec:zbehavior}, we show that the jump is
				given by:
				\beq
					\Delta Z = \left.Z\right|^{+\infty}_{-\infty} = -\frac{8 \pi \mu r Y_{lm}^{*}}{(\lambda + 1)(\lambda r + 3)}
				\eeq
				Substituting this into (\ref{eq:imp-e1}), we find an impulse of:
				\beq
					\label{eq:imp-e2} \bar f_{lm}^{\rm (e)} = \frac{16\pi \mu Y_{lm}^{*(0)}}{\lambda + 1}
				\eeq
				Note that this scales as $l^{-2} Y_{lm}^{*(0)}$ for large $l$.
			
			\subsubsection{Odd Parity, $l \geq 2$} \label{sec:impulseodd}
				Here, we integrate $f_{lm}^{\rm (o)}$ over time with the help of (\ref{eq:flmo}), to obtain:
				\beq
					\label{eq:imp-o1} \bar f_{lm}^{\rm (o)} = 8 \left.Q\right|^{+\infty}_{-\infty} - 2 \int_{-\infty}^\infty{Q(v')dv'}
				\eeq
				Unlike the Zerilli function, the Regge-Wheeler function approaches zero as $v \rightarrow \pm\infty$, so the $Q$ term in (\ref{eq:imp-o1}) drops out.  The time integral of $Q$, however, does not vanish, and as we show
				in Appendix \ref{sec:qbehavior}, this term scales as $l^{-4} Y_{lm,\theta}^{*(0)}$.  Thus, for large $l$,
				\beq
					\label{eq:imp-o2} \bar f_{lm}^{\rm (o)} \sim l^{-4} Y_{lm,\theta}^{*(0)} \sim l^{-3} Y_{lm}^{*(0)} % Is this too sloppy?
				\eeq
				We need not compute the precise form.  What matters is that the odd term (\ref{eq:imp-o2}) scales as $l^{-3}$, while the even term scales as $l^{-2}$.  This means that for large values of $l$, the odd terms may be
				neglected and only the even impulse need be considered.
				
			\subsubsection{Monopole and Dipole Terms}
				\emph{The monopole term} accounts for the mass increase of the large black hole, which, to first order, is $\mu$.  By the No-Hair Theorem, we infer that on long timescales $|v| \gg 1$, it gives rise to the following
				radial perturbation:
				\beq
					\delta r \rightarrow \left\{\begin{array}{ll} 2\mu & (v \rightarrow \infty) \\ 0 & (v \rightarrow 0) \end{array}\right.
				\eeq
				Since the monopole term is isotropic, it does not give rise to angular perturbations.
			
				\emph{The odd-parity dipole term} likewise cannot be viewed as an instantaneous impulse, but instead should be treated as a constant forcing term on long timescales:
				\beq
					f_{1m}^{\rm (o)} = -\frac{1}{2}\sqrt{4\pi/3}\,m_0 L^m \Theta(v)
				\eeq
				Using Eqs.\ (\ref{eq:geo-ph}) and (\ref{eq:fph}), we see that this results in a slow rotation at a rate $\dot{\phi}(v) = \frac{1}{8} L$ (if $\vec{L}$ is chosen to point along the $z$-axis).  Recalling that the horizon
				perturbation is done along lines of constant $u$ (for which $dt = 2dv$), we see that, in the Schwarzschild frame, the rotation rate is nothing more than $\vec{\Omega}_{\rm Kerr} = \vec{L}/4$ for slowly rotating Kerr
				black holes.
				
				Unlike its odd counterpart, \emph{the even-parity dipole term} may be treated as an impulse.  Substituting (\ref{eq:dipole-h}) into (\ref{eq:flme0}), it takes the following form:
				\beq
					f_{1m}^{\rm (e)} = 8\pi\mu e^{-v/2}\Theta(v) Y_{1m}^{*(0)}
				\eeq
				On long time scales, this resembles an impulse of the form:
				\beq
					\bar{f}_{1m}^{\rm (e)} = 16\pi\mu Y_{1m}^{*(0)} = \frac{16\pi\mu Y_{1m}^{*(0)}}{\lambda+1}
				\eeq
				This agrees with Equation (\ref{eq:imp-e2}), which we calculated only for $l \geq 2$.  Thus, (\ref{eq:imp-e2}) is valid for all $l \geq 1$.
	
		\subsection{Shape of the Impulse} \label{sec:impulseshape}
			In this section, we have shown that the forcing terms can be approximated by an instantaneous impulse.  Near the caustic, this approximation is certainly valid, as we show in Appendix
			\ref{sec:approx-validity}, so the shape of the impulse has no first-order effect on the caustics; however, it will prove enlightening to consider its shape nonetheless.
		
			As we showed in Figure \ref{fig:ResearchPaper2-Fig1}, as we increase the value of $l$, the forcing term approaches the following scaled limit curve:
			\beq
				f_{lm}^{\rm (e)} \rightarrow 8\pi\mu l^{-1} Y_{lm}^{*(0)} e^{-l|v|/2}
			\eeq
			In previous sections in this paper, we approximated this as a delta-function.  Here, we choose to retain the time dependence and proceed directly to calculate the effect on the horizon generators.  Without loss of
			generality, assume that the small black hole falls into the horizon at $\theta = 0$.  Then only terms with $m = 0$ need be considered.  Restricting ourselves to the region $\theta \ll 1$, we need consider only those terms
			with large $l$
			values.  The forcing terms at $(\theta, \phi)$ will be given by:
			\begin{eqnarray}
				F^\theta & = & \frac{1}{16} \frac{\partial}{\partial\theta} \sum_l{f_{l0}^{\rm (e)}(v) Y_{l0}} \\
				F^\phi & = & \frac{1}{16} \frac{1}{\sin^2\theta} \frac{\partial}{\partial\phi} \sum_l{f_{l0}^{\rm (e)}(v) Y_{l0}} = 0
			\end{eqnarray}
			Proceeding, we simplify $F^\theta$ by noting that $Y_{l0}$ may be written as a Bessel function for $\theta \ll 1$:
			\begin{eqnarray}
				F^\theta & = & \frac{\pi\mu}{2} \frac{\partial}{\partial\theta} \sum_l{l^{-1} Y_{l0}^{*(0)} Y_{l0}(\theta,\phi) e^{-l|v|/2}} \nonumber \\
					& \sim & \frac{\mu}{4} \sum_l{l J'_0(l\theta) e^{-l|v|/2}}
			\end{eqnarray}
			We recall that the forcing term is very nearly an impulse, and therefore $F^\theta$ will only be significantly nonzero when $|v| \ll 1$.  Taking $|v| \ll 1$, we see that the term inside the sum behaves fairly smoothly
			as a function of $l$ -- that is, it does not change much if we increase $l$ to $l+1$, or decrease it to $l-1$.  Thus, we can replace the discrete sum with an integral over $l$, and evaluate the integral analytically:
			\begin{eqnarray}
				F^\theta & \rightarrow & \frac{\mu}{4} \int_0^\infty{l J'_0(l\theta) e^{-l|v|/2}dl} \nonumber \\
					& = & \frac{\mu}{4\theta^2} \int_0^\infty{e^{-(|v|/2\theta)\xi} \xi J'_0(\xi)d\xi} \nonumber \\
					& = & -\frac{\mu}{4}\frac{\theta}{(\theta^2+v^2/4)^{3/2}} \label{eq:impulseshape}
			\end{eqnarray}
			A null generator starting at $(\theta,\phi)$ can be thought of as a light ray scattering off of the small black hole with an impact parameter $b = 2\theta$.  The transverse acceleration of the ray is given by:
			\begin{eqnarray}
				a^\perp & = & 2 F^\theta \nonumber \\
					& = & -2\mu \frac{b}{(b^2 + v^2)^{3/2}} \label{eq:a-transverse}
			\end{eqnarray}
			Recall that this is the \emph{transverse acceleration} of null rays on the horizon induced by the gravitational field of the small mass $\mu$.  If we transform into what is analogous to a Fermi normal coordinate system
			centered around the geodesic of the infalling mass, the metric near the small black hole will be locally Schwarzschild.  Near the small black hole, the horizon generators become light rays propagating through the
			(locally) Schwarzschild spacetime of the normal coordinate frame.  For weakly scattered light rays in a Schwarzschild metric, it has long been known that the transverse acceleration is given by (\ref{eq:a-transverse}),
			which is twice that predicted by Newton's theory of gravitation.\cite{Eddington}
			
			As our exercise shows, the perturbation theory gives a result (\ref{eq:a-transverse}) which agrees with the intuitive result we would expect if we considered only the \emph{local behavior} of the horizon generators in the
			vicinity of the small black hole, on spatial and time scales much smaller than the large black hole's radius of curvature, so the large black hole's field does not affect the result.  Near the small black hole, the impulse
			is the same as it would be if the large black hole had been absent.

		\subsection{Results and Accuracy}
			Substituting (\ref{eq:imp-e2}) into (\ref{eq:fr}-\ref{eq:fph}), the impulse approximation gives the following forcing terms (valid for $l \geq 2$ and even-parity $l = 1$)
			\begin{eqnarray}
				\label{eq:dfr} F^r & = & -4\pi\mu \delta(v) \sum_{lm} \frac{ Y_{lm}^{* (0)} Y^{lm}}{\lambda+1}   \\
				\label{eq:dfth} F^\theta & = &\pi\mu \delta(v) \sum_{lm}\frac{  Y_{lm}^{*(0)}Y^{lm}_{,\theta} }{\lambda+1} + {\rm (odd\ terms)} \\
				\label{eq:dfph} F^\phi & = & \pi\mu\delta(v) \sum_{lm}\frac{  Y_{lm}^{*(0)}Y^{lm}_{,\phi}}{\sin^2\theta (\lambda+1)} + {\rm (odd\ terms)} 	
			\end{eqnarray}
			We have not calculated the odd-parity terms explicitly, since as we showed in Section \ref{sec:impulseodd}, in the near-caustic impulse approximation ($\theta \ll 1$), their effect is negligible compared to the even terms.
			In the far-from-caustic impulse approximation ($\theta \gg \sqrt{m}$, $|v| > 1$), the odd terms will be comparable to the even terms.
			
			The $l = 0$ and \emph{odd-parity} $l = 1$ terms are an exception to our delta-fucntion rule, since they result in a permanent change in the black hole's mass and angular momentum, respectively, and cannot be treated as
			impulses.
			
			In Appendix \ref{sec:approx-validity}, we show that for rays in the vicinity of the caustic -- that is, for $\theta \sim \sqrt{\mu}$, the \emph{relative error} due to imposing the impulse approximation is at most of order
			$O(\mu^{1/4})$.  This is done by splitting the expressions (\ref{eq:dfr}-\ref{eq:dfph}) into terms with $l \lesssim \mu^{-1/4}$ and terms with $l \gtrsim \mu^{-1/4}$,  showing that the former can be neglected up to a relative
			error $O(\mu^{1/2})$, and that the latter can be approximated as a delta function up to a relative error of at most $O(\mu^{1/4})$.  This is only an upper bound, however, and empirically, the relative error appears to be of
			order $O(\mu^{1/2})$.
			
			The radial forcing term, by contrast, is not dominated by the large-$l$ terms.  As we show in Section \ref{sec:caustics}, $F^r$ is of order $m \log m$, while the low-$l$ perturbations give a contribution of $O(m)$.  Therefore, the low-$l$ terms cannot be ignored in the radial case as they could be in the angular case.  However, from (\ref{eq:fr}-\ref{eq:fph}), we see that the \emph{derivatives} of $F^r$ can be related to the angular terms by $\nabla^a F^r = -4 F^a_{\rm (even)}$, which is well approximated as an impulse.  This identity is exact and holds for all $l$.  If we then restrict ourselves to a small patch of the horizon (say, the neighborhood of the caustic), $F^r$ will equal its large-$l$ contribution up to a time-varying but constant function $C^{(0)}$.  The impulse approximation for $F^r$ thus holds up to a \emph{constant term}:
			\beq
				\label{eq:dfr-2} F^r = -4\pi\mu \delta(v) \sum_{lm} \frac{ Y_{lm}^{* (0)} Y^{lm}}{\lambda+1} + C^{(0)}(v)
			\eeq
			The constant $C^{(0)}(v)$ depends on the time coordinate, but is independent of the $\theta$ and $\phi$ coordinates, falls off to zero as $|v| \gg 1$, and as we will show, does not affect the properties of the caustic.
			
			Far from the caustic, the deflection of any ray is very small -- $\delta\theta \ll \theta$ -- and therefore the impulse approximation will be valid at large times up to a relative error $O(\delta\theta/\theta) \ll 1$.
			However, it is important to recall that far from the caustic, the impulse approximation only holds on timescales $|v| \gg 1$, as described in Section \ref{sec:impulse-bulk}.  The actual ``impulse'' is spread out over
			a timescale $O(1)$, and therefore does not look like an impulse for timescales $v \sim O(1)$.  Also, unlike the caustic-vicinity case, in the bulk, odd-parity contributions to the impulse are not negligible.
		
	\section{Event Horizon Caustics} \label{sec:caustics}
		A remarkable feature or black hole merger spacetimes is the presence of \emph{caustics} -- regions where the null generators on the event horizon cross each other, giving rise to discontinuous kinks in the event horizon's
		geometry.  While the \emph{event horizon} may be sharp and pointed in the vicinity of a caustic, the \emph{metric} remains smooth; therefore perturbation theory can be applied to study the structure of these caustics.  The
		impulse approximation in the above section simplifies matters greatly, allowing the structure of the caustic to be determined analytically as a function of the mass ratio.
		
		\subsection{Dynamics of Generators that form Caustics}
			Let us define coordinates $x^a = (\theta, \phi)$ on the unit sphere.  Using this notation, we may recast Equations (\ref{eq:geo-r}-\ref{eq:geo-ph}) as:
			\beq
				\left(\frac{d}{dv} - \frac{1}{4}\right)\frac{dx^a}{dv} = F^a(x^b, v)
			\eeq
			Noting that in the orthonormal basis of spherical harmonics $2(\lambda+1) = -\nabla^2$ (the Laplacian being taken on the unit sphere) and $Y_{lm}^{*(0)} = \int{d\Omega Y_{lm}^* \delta^{(2)}(\Omega)}$, we may simplify 
			Equations (\ref{eq:dfth}-\ref{eq:dfph}) and (\ref{eq:dfr-2}) to give:
			\begin{eqnarray}
				\label{eq:cfr} F^r & = & -4\pi \mu (-2(\nabla^2)^{-1}) \delta^{(2)}(\Omega) \delta(v) + C^{(0)}(v) \\
				\label{eq:cfvec} F^a & = & \pi \mu \nabla^a(-2(\nabla^2)^{-1}) \delta^{(2)}(\Omega) \delta(v)
			\end{eqnarray}
			from which it is clear that
			\beq
				\label{eq:fr-fvec} F^a = -\frac{1}{4} \nabla^a F^r
			\eeq

			\begin{figure}[tbp]
				\centering
					\includegraphics[width=1.00\columnwidth]{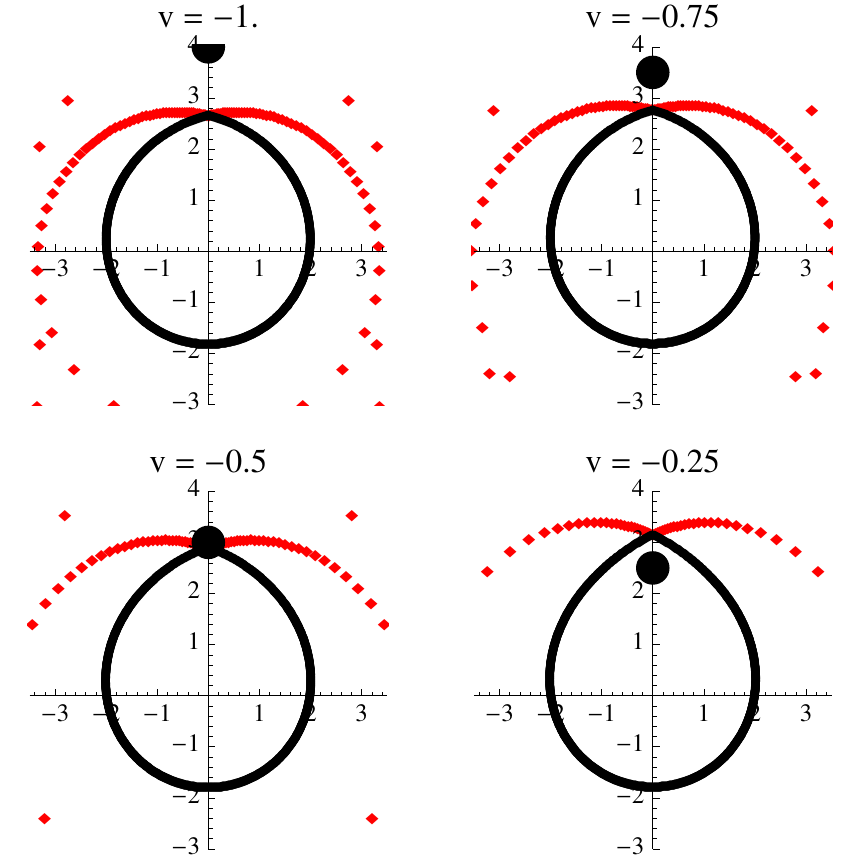}
				\caption{Event horizon and caustic of a black hole merger with $m = 0.15$, plotted in an ingoing coordinate system at times $v = -1$, $-0.75$, $-0.5$, and $-0.25$ respectively.  Black dots indicate null generators on the
					horizon; red dots indicate rays which have yet to enter the horizon via the caustic.  The large black dot is the infalling black hole. }
				\label{fig:Fig1a}
			\end{figure}

			Therefore, finding the forcing terms boils down to solving Poisson's Equation for a point source $\mu \delta^{(2)}(\Omega)$.
			Noting that the ``potential''
			\beq
				\Psi \equiv \frac{1}{4\pi}\ln\left[1-\cos\theta\right]
			\eeq
			satisfies $\nabla^2\Psi = \delta^{(2)}(\Omega)$ (at least for $l \geq 2$, where the Zerilli formalism if valid), the forcing terms take the following form:
			\begin{eqnarray}
				F^r & = & 8\pi \mu \Psi \delta(v) + C^{(0)}(v) \nonumber \\
				\label{eq:cfr2} & = & 2\mu \ln\left[1-\cos\theta\right] \delta(v) + C^{(0)}(v) \\
				\label{eq:cfvec2} F^a & = & -2\pi \mu \nabla^a \Psi \delta(v) = -\frac{1}{2} \mu \frac{1 + \cos\theta}{\sin\theta}\delta(v)\hat{e}_{\theta}^a 
			\end{eqnarray}
			($\hat{e}_{\theta}^a$ is the $\theta$ unit vector on the unit sphere).
			
			In this equation, we have neglected the $l = 0$ and odd-parity $l = 1$ terms, which do not function as impulses.  That aside, (\ref{eq:cfr2}-\ref{eq:cfvec2}) give a general expression for the forcing terms and may be used
			both in the caustic vicinity, and in the bulk, of the event horizon, under the conditions spelled out in Section \ref{sec:delta-function}.  In this section, we are interested in the caustic structure and accordingly treat
			the case $\theta \ll 1$, where the $l = 0$ and $l = 1$ terms play an insignificant role.  In this regime, (\ref{eq:geo-r}-\ref{eq:geo-ph}) take the following form:

			\begin{eqnarray}
				\left(\frac{d}{dv} - \frac{1}{4}\right) \delta r & = & 4\mu (\log(\theta) + C) \delta(v) \nonumber \\
				\label{eq:caustic-r} & & + C^{(0)}(v) - 2\left(\frac{d\theta}{dv}\right)^2 \\
				\label{eq:caustic-th} \left(\frac{d}{dv} - \frac{1}{4}\right) \frac{d}{dv} \theta & = & -\frac{\mu}{\theta} \delta(v) \\
				\label{eq:caustic-ph} \phi & = & \mbox{const}
			\end{eqnarray}
			where $C = -\frac{1}{2}\log(2)$ is a constant.
			
			To illustrate the effect of (\ref{eq:caustic-r}-\ref{eq:caustic-ph}) on the null generators, let us trace out the path of a given generator.  Since the generator is a part of the \emph{future} event horizon, we ``start'' it
			at future null infinity ($v = \infty$) and propagate it ``backward'' in time.  Let $r_+$, $\theta_+$, and $\phi_+$ denote the values of $r$, $\theta$, and $\phi$ the ray ``starts'' with at $v \rightarrow +\infty$.  Likewise
			denote $r_-$, $\theta_-$, and $\phi_-$ as the values the ray ``ends'' with at $v \rightarrow -\infty$.  Equations (\ref{eq:caustic-r}-\ref{eq:caustic-ph}) are straightforward to solve; for $v > 0$ (after the plunge), 
			we have:
			\begin{eqnarray}
				r(v) & = & r_+ + c(v) \\
				\theta(v) & = & \theta_+ \\
				\phi(v) & = & \phi_+
			\end{eqnarray}
			Prior to the merger, the solution takes the form:
			\begin{figure}[tbp]
				\centering
					\includegraphics[width=\columnwidth]{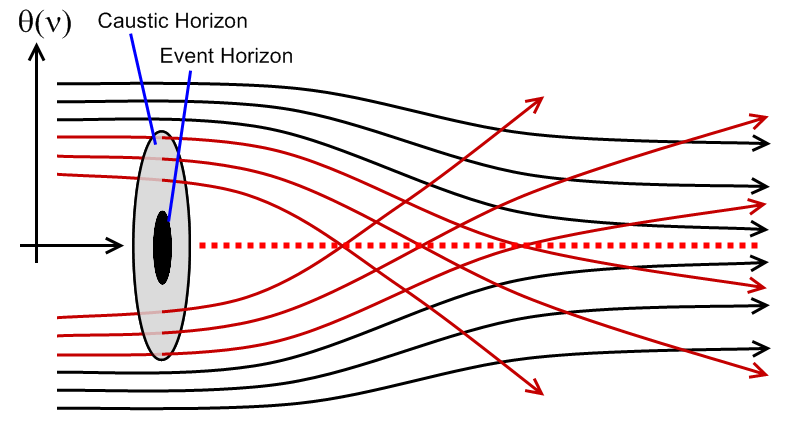}
				\caption{Null event horizon generators, traced back in time.  The rays inside the caustic horizon cross each other and exit the event horizon, forming caustics.  The rays outside the caustic horizon always remain on the
					event horizon.}
				\label{fig:Fig2}
			\end{figure}

			\begin{eqnarray}
				r(v) & = & r_- + \kappa e^{v/4} \nonumber \\
				\label{eq:caustic-r2} & & - (8\mu^2/\theta_+^2)(e^{v/2}-e^{v/4}) + c(v) \\
				\label{eq:caustic-th2} \theta(v) & = & \theta_- + (\theta_+ - \theta_-) e^{v/4} \\
				\phi(v) & = & \phi_+
			\end{eqnarray}
			where

%			\begin{eqnarray}
%				\label{eq:caustic-r2} r(v) & = & \left\{\begin{array}{ll} r_+ & (v > 0) \\ r_- + \kappa e^{v/4} - (8m^2/\theta_+^2)(e^{v/2}-e^{v/4}) + c(v) & (v < 0) \end{array}\right. \\
%				\label{eq:caustic-th2} \theta(v) & = & \left\{\begin{array}{ll} \theta_+ & (v > 0) \\ \theta_- + (\theta_+ - \theta_-) e^{v/4} & (v < 0) \end{array}\right. \\
%				\phi(v) & = & \phi_+
%			\end{eqnarray}
%			where
		
			\begin{eqnarray}
			 	r_- & = & r_+ = 2 \\
				\kappa & = & -4 \mu\left[\log(\theta_+) + C\right] \\
				\label{eq:thm} \theta_- & = & \theta_+ - \frac{4\mu}{\theta_+}
			\end{eqnarray}
			and $c(v)$ is defined so that
			\beq
				\label{eq:cv-def} c'(v) - \frac{1}{4}c(v) = C^{(0)}(v)
			\eeq
			These solutions are visualized in Figure \ref{fig:Fig1a}.  Note that, as expected, rays enter the event horizon through the caustic at $\theta = 0$.  The 2-dimensional time slices of the horizon develop kinks at the 
			caustic, in agreement with the results of  previous literature.\cite{HeadOn}  Note further that the two constants in this result, $C$ and $c(v)$, do not affect the internal properties of the caustic, but merely shift it
			in the $r$-direction.
				
			For an infalling point mass, equation (\ref{eq:caustic-th}) tells us that caustics will always form for rays of small enough $\theta_+$.  Naturally, we are inclined to ask:  When traced back to $v = -\infty$, what does
			the set of rays that form caustics look like?  It is fairly easy to see from (\ref{eq:thm}) that a ray will form a caustic if:
			\beq
				\theta_+ < 2 \mu^{1/2}
			\eeq
			Define the \emph{caustic horizon} as the set of rays with $\theta_+ = \theta_c \equiv 2\mu^{1/2}$ (see Fig. \ref{fig:Fig2}).  When traced back in time, the rays inside the caustic horizon form caustics and leave the event horizon, while rays outside
			the caustic horizon remain on the event horizon forever.  Equivalently, rays inside the caustic horizon originate outside the event horizon and enter it through the caustic, while rays outside the caustic horizon originate
			on the event horizon.
			
		\subsection{Properties of the Caustic}

			\begin{figure}[tbp]
				\centering
					\includegraphics[width=0.80\columnwidth]{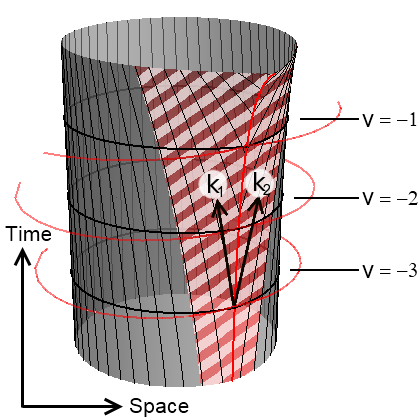}
					\caption{The event horizon of the large black hole as a surface in spacetime.  The small black hole has a mass $m = 0.15$.  Null generators are shown as black lines, and the region spanned by the generators entering
					through the caustic is shaded red.}
				\label{fig:R2-Fig9}
			\end{figure}
		
			Globally, the caustic is a spacelike line which lies on the future horizon of the black hole.  This line can be associated with a function $\delta r(v)$, defined by solving Eqs.\ (\ref{eq:caustic-r2}-\ref{eq:caustic-th2})
			subject to the constraint $\theta = 0$.  We find:
			\beq \label{eq:global-dr}
				\delta r_c = -4\mu\left[\log\sqrt{4\mu(1-e^{v/4})} + \left(C - \frac{1}{2}\right)\right] e^{v/4} + c(v)
			\eeq
			The caustic does {\it not} lie along a null generator; it is spacelike and its length is a well-defined and invariant quantity.  Taking the line element along the path of the caustic, we find:
			\begin{eqnarray}
				ds^2 & = & \left[-(1-2/r_c) + \delta g_{vv}\right]dv^2 + 2dv\,dr_c \nonumber \\
				& = & \left[\mu\frac{e^{v/2}}{1-e^{v/4}} + 2C^{(0)}(v) - 2F^r\right] dv^2 \nonumber \\
				\label{eq:dsdt} & = & \mu\frac{e^{v/2}}{1-e^{v/4}} dv^2
			\end{eqnarray}
			Once again, we find that the measurable quantities of the caustic do not depend on the ``constants'' $C$ or $C^{(0)}(v)$.  The result (\ref{eq:dsdt}) integrates to a total length
			\beq
				L_c = \int_{-\infty}^{0}{\sqrt{\mu} \frac{e^{v/4}}{\sqrt{1-e^{v/4}}} dv} = 8\sqrt{\mu}
			\eeq
			Even though the caustic stretches back to $v = -\infty$, it has a finite invariant length.
			
			\begin{figure}[tbp]
				\centering
					\includegraphics[width=0.80\columnwidth]{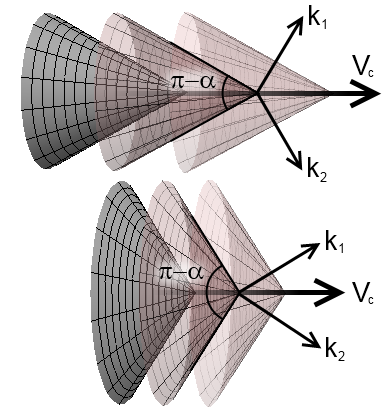}
				\caption{Viewed as a 2-surface in 3-space, the horizon near the caustic looks like a cone of angle $\pi - \alpha$.  This is shown for a small cone angle (top) and a larger cone angle (bottom), both cones propagating to
				  the right as time progresses (three time slices of the horizon are shown in the figures).  Like a Cherenkov cone, the speed $V_c > c$ increases as the cone angle decreases.}
				\label{fig:R2-Fig10}
			\end{figure}
			
			Locally -- i.e. in a local Lorentz frame which includes the caustic -- the horizon around the caustic resembles a cone of angle $\pi - \alpha$ propagating outward at a superluminal velocity $V_c$, as illustrated in Figure
			\ref{fig:R2-Fig10}.  At least two null rays ($k_1$ and $k_2$, whose spatial components are shown in the figure; see also Figure \ref{fig:R2-Fig9}) tangent to the horizon meet at the caustic.  As we showed previously, our
			caustic has a conical shape, so an entire cone of null rays meet at the caustic.  To keep things simple, though, we only consider two such rays, chosen to be tangent to opposite sides of the cone.
			
			From geometric considerations, we can relate the cone angle to the speed of the caustic.  We find:
			\beq
				V_c \equiv \frac{dx}{dt} = \sec(\alpha/2)
			\eeq
			where $x$ and $t$ are spatial and time coordinates in the local Lorentz frame.  Thus, a very sharp cone propagates very quickly, while a blunter cone will propagate slower, but still superluminally.  The speed $V_c$ is in
			turn related to the invariant distance traced out per unit time:
			\beq \label{eq:dsdt-angle}
				\frac{ds}{dt} = \sqrt{V_c^2 - 1} = \tan(\alpha/2)
			\eeq
			A third local property of the caustic is the \emph{deviation between generators} at the caustic.  If, in a given time slicing, we normalize the horizon generators so that $k_i^t = 1$, we can define a quantity $\psi_c$ as
			follows:
			\beq
				\psi_c = \frac{1}{2}\max_{k_1,k_2}\left[\left|k_1 - k_2\right|\right]
			\eeq
			The quantity is maximized over all generators $(k_1,k_2)$ tangent at the caustic, the maximum being obtained when the vectors point along opposite sides of the cone.  Again, elementary geometric considerations relate
			$\psi_c$ to the cone angle as follows:
			\beq \label{eq:psi-angle}
				\psi_c = \sin(\alpha/2)
			\eeq
			For $\alpha \ll 1$, we therefore have $\psi_c \approx ds/dt$.
			
			Choosing $v = t + r_*$ as our time parameter, we can foliate the spacetime into three-dimensional slices.  Using this slicing, we can derive the invariant distance per unit time using Equation (\ref{eq:dsdt}):
			\beq
				ds/dv = \sqrt{\mu} \frac{e^{v/4}}{\sqrt{1-e^{v/4}}}
			\eeq
			Likewise, solving (\ref{eq:caustic-th2}) for $\dot{\theta}(v)$, holding $\theta(v) = 0$, we obtain a null generator $k^\mu = (1, O(\mu), \dot{\theta}, 0)$ at the caustic, where 
			$\dot{\theta} = \sqrt{\mu} e^{v/4}/\sqrt{4(1-e^{v/4})}$.  Using the metric at the horizon, the deviation between generators is then found to be:
			\beq
				\psi_c = \sqrt{\mu} \frac{e^{v/4}}{\sqrt{1-e^{v/4}}}
			\eeq
			Both of these properties, derived independently from each other, point to a cone angle of
			\beq
				\pi - \alpha,\ \ \ \ \alpha = 2\sqrt{\mu} \frac{e^{v/4}}{\sqrt{1-e^{v/4}}}
			\eeq
			and are in agreement with each other.  For small $\mu$, the caustic angle is very nearly equal to $\pi$ for all reasonable times $|v| > \mu$ far from the small hole's event horizon (where the perturbation theory is valid).
			Only very near the merger -- i.e. in the regime $|v| < \mu$ that the perturbation theory cannot probe -- does the caustic angle deviate significantly from $\pi$.
			
			Of course, these results should be taken with a grain of salt.  While the caustic angle does not depend on the spatial coordinates we choose, it \emph{does} depend on the time slicing.  By a clever choice of coordinates,
			one may make $\alpha$ take any value one desires.  However, relations (\ref{eq:dsdt-angle}) and (\ref{eq:psi-angle}) hold irrespective of the coordinate choice, allowing the caustic angle to be related to the invariant
			distance per unit time and the deviation between generators in any time slicing.  Moreover, the integrated caustic length $L_c = 8\sqrt{\mu}$ is a slicing-invariant quantity as well.
			
			It is worth emphasizing that the results derived in this section are independent of the particulars of the plunge.  We therefore say that the caustic structure is \emph{universal} -- it depends on the mass-energy of
			the infalling black hole, and not on anything else.				
		
			Before we end this section, a few consistency checks are in order.  First the perturbation theory breaks down when the field of the point mass is strong -- i.e. near the small black hole's event horizon.
			Since the size of the event horizon is proportional to $\mu$, our results are only valid for $\theta_+ \gg \theta_{EH} \sim \mu$.  Note, however, that the caustic horizon scales as $\theta_c \sim \mu^{1/2}$, so that in the
			extreme mass-ratio limit ($\mu \ll 1$), $\theta_c \gg \theta_{EH}$.  Put in words, for a small infalling mass, the event horizon is \emph{tiny} compared to the caustic horizon, and the field of the infalling mass is weak
			for rays of $\theta_+ \sim \theta_c$.  Most caustic-forming rays, therefore, can be accurately described using the methods of this section.
		
			Second, in order for geodesic perturbation theory to be valid, we must require that $\delta r \ll 1$.  At first glance, this appears to fail by virtue of the logarithmic term in (\ref{eq:caustic-r2}).  However, a closer
			inspection reveals that in the limit of small $\mu$, this term is of order unity only for rays of $\theta_+ = O(e^{-1/\mu}) \ll O(\mu)$.  As explained above, the gravitational field of $m$ is strong for $\theta_+ = O(\mu)$,
			and the theory breaks down anyway.  In the region where the theory \emph{does} apply ($\theta_+ \gg \mu$), the radial perturbation is small everywhere.

	\section{Black Hole Area Increase} \label{sec:area}
		The surface area of a black hole of mass $M$ is $16\pi M^2$.  For an infinitesimal mass increase, the surface area increases by $\Delta A = 32\pi M \Delta M$.  Thus, for the case in question -- a point mass $\mu \ll 1$
		falling into a black hole of mass $M = 1$, we expect the black hole's event horizon to grow by $\Delta A = 32\pi \mu$.  Naturally, we are inclined to ask: how much of this area increase is due to rays which enter the
		horizon through the caustic, and how much of it is due to the expansion of the horizon itself?
		
		\subsection{General Principles}	\label{sec:genprinc}	
\begin{figure}[tbp]
				\centering
					\includegraphics[width=1.00\columnwidth]{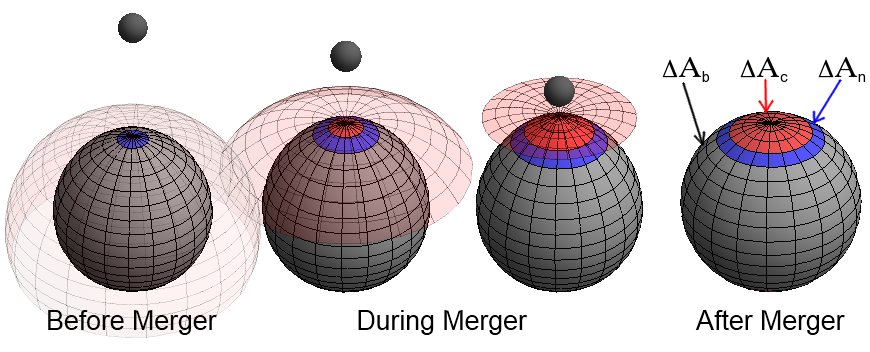}
				\caption{As black holes merge, the surface area of the large hole's horizon increases.  This is due to rays entering the horizon through the caustic ($\Delta A_c$, shown in red), expnsion of rays near the caustic 
				($\Delta A_n$, blue), and expansion in the bulk of the horizon ($\Delta A_b$, gray).}
				\label{fig:Fig3}
			\end{figure}

			Without loss of generality, assume the small black hole reaches the horizon at the $+z$ axis -- that is, with $\theta = 0$.
			We can then write the horizon area increase as a sum of three parts: $\Delta A = \Delta A_c + \Delta A_n + \Delta A_b$.  Here $\Delta A_c$ refers to the area increase due to rays entering through the caustic and $\Delta
			A_n$ refers to the expansion of area elements in the ``neighborhood'' of the caustic -- i.e. for $\theta \ll 1$.  The last term, $\Delta A_b$, refers to the increase due to the expansion of the bulk of the horizon.  Figure
			\ref{fig:Fig3} illustrates our point.
		
			A number of general principles allow one to solve the area increase problem with minimal effort.  First, as mentioned above, the total area increase must be $32\pi \mu$.  Second, infinitesimal area elements can expand, but
			they cannot contract.\cite{HawkingPaper}  Third, area elements in the bulk of the horizon are minimally affected by the point mass's gravitational field, and therefore neither expand nor contract.
		
		\subsection{Detailed Calculation}
			The purpose of this subsection is to verify the points made above with explicit calculations of the three area contributions.

			\subsubsection{Area due to generators entering through the caustic}
			
			Recalling that rays for which $\theta_+ < \theta_c = 2\sqrt{\mu}$ enter through the caustic, the area increase due to the caustic is simply equal to the area spanned by these generators on the final horizon, which is equal
			to
			\beq
				\Delta A_c = \pi (2 \theta_c)^2 = 16\pi \mu
			\eeq
			This is half of the total area increase.  

			\subsubsection{Area increase in caustic vicinity}
			
				Consider, in the neighborhood of the caustic, an annulus $\theta \in \left[\theta_+, \theta_+ + d\theta_+\right]$ at future infinity ($v = +\infty$).  Tracing these rays back to past infinity, we end up with a new
				annulus defined by:
				\begin{eqnarray}
					\theta & \in & \left[\theta_-, \theta_- + d\theta_-\right] \\
					& = & \left[\theta_+ - \delta\theta(\theta_+), \theta_+ + d\theta_+ - \delta\theta(\theta_+ + d\theta_+)\right]
				\end{eqnarray}
				where $\delta\theta(\theta_+)$ is the $\theta$ deflection of the ray due to the small hole's gravitational field.  Assuming that the annulus is thin, i.e. $d\theta_+ \ll \theta_+$, we find the new annulus has radius and
				width:
				\begin{eqnarray}
					\theta_- & = & \theta_+ - \delta\theta(\theta_+) \\
					d\theta_- & = & d\theta_+\left[1 - \frac{\partial \delta\theta}{\partial\theta}\right]_{\theta_+}
				\end{eqnarray}
				At past infinity, the area of the annulus is $A_- = 8\pi \sin(\theta_-) d\theta_-$.  After the merger, the area grows to $A_+ = 2(2+\delta r)^2\pi \sin(\theta_+) d\theta_+$.  The infinitesimal area increase is given by:
				\begin{eqnarray}
					\delta A & = & A_+ - A_- \nonumber \\
					& = & 8\pi \left[\sin(\theta_+)d\theta_+ - \sin(\theta_-)d\theta_- + \delta r \sin(\theta_+)d\theta_+\right] \nonumber \\
					\label{eq:a-increase} 
				\end{eqnarray}
				This is a general formula that applies equally to the caustic neighborhood and the bulk.
				
				Specializing to the caustic neighborhood, we set $\theta_+ \ll 1$.  Equation (\ref{eq:a-increase}) describes the area increase as due to two components -- radial deformation, and angular deformation.  The radial
				deformation is trivial -- at future infinity, the radius of the black hole increases by $\delta r = 2\mu$.  In Section \ref{sec:caustics}, we calculated the angular deformation to be: $\delta\theta(\theta_+) =
				4\mu/\theta_+$.  Applying these substitutions, we arrive at the following area increase:
%				\beq
%					\label{eq:a-increase-n} \delta A = \left[128\pi \mu^2 \frac{1}{\theta_+^3} + 16\pi\mu\theta_+\right]d\theta_+
%				\eeq
				\beq
					\label{eq:a-increase-n} \delta A = \bigl[\underbrace{128\pi \mu^2 \theta_+^{-3}}_{\rm angular} + \underbrace{16\pi\mu\theta_+}_{\rm radial}\bigr]d\theta_+
				\eeq
				The angular part dies off quickly with increasing $\theta$ and therefore only contributes in the neighborhood of the caustic.  Ostensibly, it appears to be a second-order effect in $m$; however, when we integrate it over
				the neighborhood, we find an total area increase of:
				\beq
					\label{eq:a-increase-n2} \int_{\theta_c}^{\theta_{max}}{128\pi \mu^2 \frac{d\theta_+}{\theta_+^3}} \rightarrow 16\pi \mu
				\eeq
				where $\theta_c \ll \theta_{\rm max} \ll 1$ so that we only integrate over the neighborhood of the caustic.
			
				The neighborhood of the caustic has an area $A_n \sim \theta_{\rm max}^2 \ll 1$.  The radial term in (\ref{eq:a-increase-n}) will increase the area by an amount $O(\mu A_n) \ll O(\mu)$, which is much less than the 
				area increase computed in (\ref{eq:a-increase-n2}).  Therefore, to leading order in $m$, the area increase in the caustic neighborhood is:
				\beq
					\Delta A_n = 16\pi \mu
				\eeq
				This contributes the other half to the total increase $32\pi \mu$.

%\textcolor{red}{Here we need a better reasoning for what is $A_+$ and what is $A_-$.  Why are we not using $r$ deformations?  The answer must be: $\delta r$  contribution for these generators are much less important.  What's the real integration upper bound?  Will that bound be Ok for impulse approximation?  The answer should still be yes, because upper bound only needs to be a large number times the lower bound for the integral to converge, yet a large number times the lower bound is still considered small enough for impulse approximation.  Please, be explicit!!}

\subsubsection{Area increase in the bulk}
			
			Next, we turn to the bulk -- that is, generators with $\theta > \theta_{\rm max}$, with $\theta_{\rm max}$ defined as above.  Like near the caustic, an area element in the bulk will expand or contract due
			to two factors: angular deformation and radial deformation.  The extent to which it does so is governed by (\ref{eq:a-increase}).  In the bulk, however, rays are weakly deflected, i.e. $\delta\theta \ll \theta_+$.  This
			allows us to simplify (\ref{eq:a-increase}) to the following form:
			
%			\textcolor{red}{What is the bulk, exactly?  Did we really define this?}

			\beq
				\delta A = A \left[\delta r + \cot\theta \delta\theta + \frac{\partial \delta\theta}{\partial\theta}\right]
			\eeq
			(Here $A$ is the size of an infinitesimal area element.  It does not matter whether we use $A_+$ or $A_-$, since in the bulk of the horizon, $A_+ = A_-$ up to a relative error $O(\mu)$). %, and we are only concerned with the leading-order area increase here).
			
			In the bulk, furthermore, the odd-parity perturbations cannot be neglected as they may in the caustic vicinity.  Nevertheless, their effect on the horizon area is immaterial.  A quick glance at Equations
			(\ref{eq:fr}-\ref{eq:fph}) reveals that the forcing terms for the odd-parity perturbations are solenoidal in nature -- they can shear area elements on the horizon, but they cannot expand them.  The even-parity perturbations
			give rise to \emph{expansion}, while the odd-parity terms give rise to \emph{shear}.  It is the expansion of area elements that chiefly concerns us here. 
			
			As above, we set $\delta r = 2\mu$; hence, $\delta r$ has no angular dependence.  In Part \ref{sec:genprinc}, we asserted that the infinitesimal area elements in the bulk neither expand nor contract -- i.e. that 
			$\delta A = 0$ everywhere in the bulk.  This gives a first-order ODE for $\delta\theta(\theta)$.  Solving subject to a continuity constraint at $\theta = \pi$, we expect an angular deformation of the form:
			\beq
				\label{eq:a-bulk} \delta\theta = 2\mu \frac{1 + \cos\theta}{\sin\theta} + \mbox{(odd terms)}
			\eeq
			We can obtain precisely the same result by invoking the impulse approximation, applying the forcing terms in Equations (\ref{eq:cfr2}-\ref{eq:cfvec2}) to the bulk, and adding the odd-parity perturbations (which do not
			affect the area increase).

% Deleted discussion that has been incorporated into other parts of the paper.
				
	\section{Caustic Formation with Strings} \label{sec:strings}
	
%	\textcolor{red}{We should change $\vec x$ etc.\ into $x^a$ etc.  I'm very uncomfortable with ``vector subtraction'' used in the following section ...}
	
		In the preceding sections, we have constrained our analysis to the infall of point particles in the extreme mass-ratio case.  For point particles, we found that the event horizon develops a caustic and that the caustic
		accounts for 50\% of the black hole area increase.  On the other hand, it should be eminently clear that extended objects \emph{larger} than the caustic horizon will not form caustics when they plunge into the black hole.
		Naturally, one is inclined to ask: what about objects extended in one or two dimensions?  In this section, we study the infall of such objects and conclude that, under certain circumstances, they do form caustics, but that
		such caustics account for 100\%, not 50\%, of the black hole area increase.
%		\subsection{Dynamics of Generators and Area}
			\subsection{Parallel Strings}
				Recall our analysis for an infalling point particle in Section \ref{sec:caustics}.  In the vicinity of the caustic horizon, $\theta \ll 1$ and the event horizon's geometry is essentially Euclidean.  We may therefore write
				$\theta$ and $\phi$ in terms of Euclidean coordinates $x^a = (x_1, x_2)$ and recast Eqs. (\ref{eq:caustic-r}-\ref{eq:caustic-ph}) as:
				\begin{eqnarray}
					\left(\frac{d}{dv} - \frac{1}{4}\right) \delta r & = & F^r(x^a, v) + C^{(0)}(v) - 2(\dot{x}_1^2 + \dot{x}_2^2)  \nonumber \\
					& = & 4\mu \left(\log(x) + C\right) \delta(v) + C^{(0)}(v) \nonumber \\
					\label{eq:geo-r1} & & - 2(\dot{x}_1^2 + \dot{x}_2^2) \\
					\label{eq:geo-x1} \left(\frac{d}{dv} - \frac{1}{4}\right)\frac{d}{dv} x^a & = & F^a(x^b, v) = -\frac{\mu}{x} \hat{x}^a \ \delta(v)
				\end{eqnarray}
				where $x = ||x^a||$ and $\hat{x}$ is the normalized vector.
				
				Let us spread out the point mass into a string parallel to the event horizon.  Since we are working to first order in the perturbation, our angular equations are linear, and thus obey the superposition principle.  
				(The radial equation is not linear, but once we have $\dot{x}^a$, it is easily solved).  We can therefore rewrite the forcing term in (\ref{eq:geo-x1}) as:
			\begin{figure}[tbp]
				\centering
					\includegraphics[width=0.75\columnwidth]{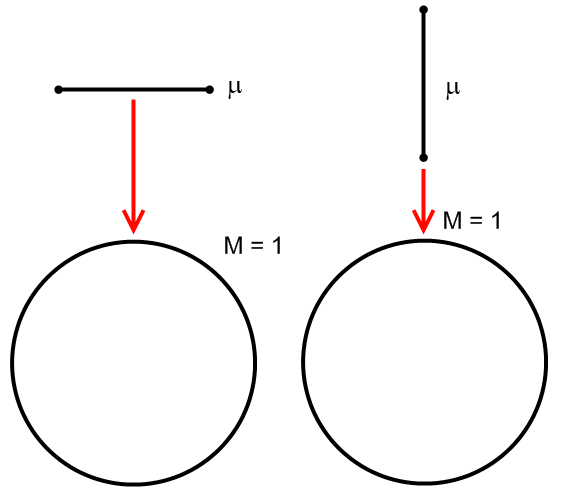}
				\caption{A diagram of two ``strings'' falling into a black hole, one parallel to the horizon and the other orthogonal.  In reality, most objects would fall in at oblique angles.}
				\label{fig:Fig5}
			\end{figure}
				\beq
					\label{eq:geo-forcing} F^a(x^b, v) = \int{-\frac{d\mu}{q} \hat{q}^a} \delta(v)
				\eeq
				where $q^a = x^a - {x^a}'$ is a vector pointing from the mass element $d\mu$ to the observer at point $x^a$, and the vectors with hats ($\hat{x}^a$, $\hat{q}^a$) represent normalized unit vectors.  Solving the equations
				of motion prior to merger, we find that the displacement of any ray due to the string will be given by:
				\begin{eqnarray}
					\delta r & = & e^{v/4} \Delta r + c_{\rm str}(v) - \frac{1}{2}(\Delta x^a)^2 (e^{v/2} - e^{v/4}) \\
					x^a & = & x_+^a + \Delta x^a (e^{v/4} - 1)
				\end{eqnarray}
				where
				\begin{eqnarray}
					\Delta r & = & -\int{4\log(q)d\mu} - 4\mu C \\
					\Delta x^a & = & \int{4\frac{d\mu}{q} \hat{q}^a}
				\end{eqnarray}
				and $c_{\rm str}(v)$, defined in the same way as (\ref{eq:cv-def}), merely shifts the caustic in the $r$-direction and has no effect on its internal structure.
				
				Both $\Delta r$ and $\Delta x^a$ obey the superposition principle, and can likewise be related to the structure of the string by analogues of Gauss's Law and Poisson's Equation:
				\beq
					-\partial_a \partial^a \Delta r = \partial_a \Delta x^a = 8\pi \rho
				\eeq
				where $\rho$ is the amount of mass which falls through the horizon per unit area in the $(x^1,x^2)$ coordinates.  The integral $\int{\rho\,dx_1dx_2}$ would then correspond to the enclosed infalling mass.

\begin{figure}[tbp]
					\centering
						\includegraphics[width=1.00\columnwidth]{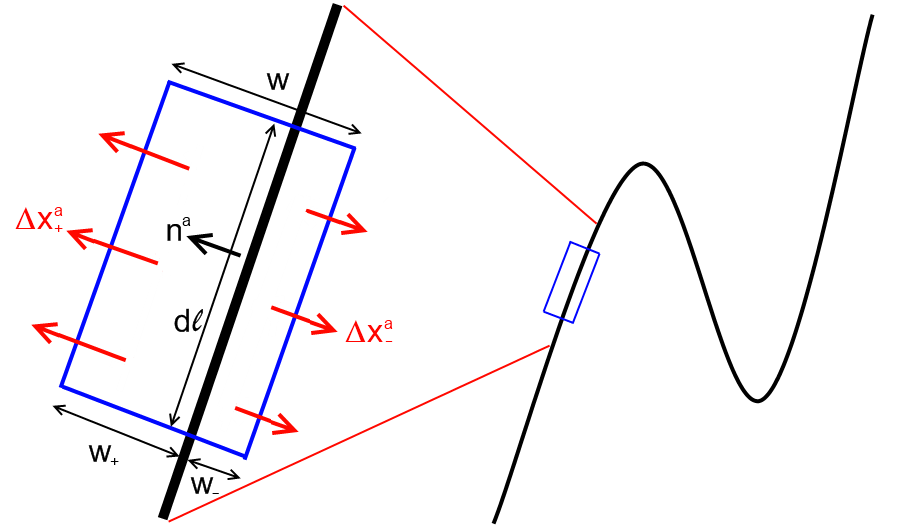}
					\caption{A Gaussian surface chosen around a string falling into the event horizon.}
					\label{fig:Fig6}
				\end{figure}

				Now consider an arbitrary string falling in parallel to the horizon.  Draw a Gaussian surface as per Figure \ref{fig:Fig6} around any small line element of the string.  A simple application of Gauss's Law allows one to
				relate the displacement vector $\Delta x^a$ on one side of the string to its value on the other side:
				\beq
					\Delta x^a \Bigr\rvert^{+}_{-} = 8\pi \Lambda \hat{n}^a
				\eeq
				The displacement quantities $\Delta r$ and $\Delta x^a$ then take the following form near the string:
				\begin{eqnarray}
					\label{eq:str-dr} \Delta r & = & A - B_a x^a - 4\pi \Lambda \lvert x\rvert \\
					\label{eq:str-dx} \Delta x^a & = & B^a + 4\pi\Lambda\,{\rm sgn}(x) \hat{n}^a
				\end{eqnarray}
				where $\Lambda$ is the linear mass density of the string and $A$ and $B^a$ are constants of integration.  Note, critically, that unlike in the case of the point particle, the displacement field $\Delta x^a$ is roughly
				constant in the vicinity of the string; it does not depend on the distance between the string and the observer.  This displacement will cause the horizon generators near the string to converge towards it, forming a
				one-dimensional, linelike caustic.
				
				Consider again the Gaussian surface in Figure \ref{fig:Fig6}.  The widths $w_+$ and $w_-$ on each side of the caustic are fixed so that $\Delta r$ takes the same value at both edges of the Gaussian surface.  Since they
				have the same $\Delta r$ value, the rays at the edges of the surface will meet if and only if they cross in angular coordinates.  This will happen if the total width $w$ is less than $8\pi\Lambda$, the ``jump'' in
				$\Delta x^a$ across the string.  Therefore, the generators entering the horizon through the caustic form a small strip, containing the string, of width $w = 8\pi\Lambda$.  Integrating over the whole string, we find the
				total area increase due to the caustic:
				\beq
					\Delta A_c = 4(8\pi \mu) = 32\pi \mu
				\eeq
				Thus, the linelike caustic of the string is responsible for 100\% of the increase in black hole area.  This contrasts sharply with the 50\% figure found for point particles.  As we derived in Equation (\ref{eq:str-dx}),
				the displacement field $\Delta x^a$ is approximately a constant in the vicinity of the caustic, as opposed to the $1/\theta$ dependence found for the point mass case.  A constant displacement field makes it impossible for
				area elements near the caustic to expand (except at the edges of the string, which contribute negligibly to the area).  By our reasoning in Section \ref{sec:area}, none of the area increase comes from the bulk to first
				order in $\mu$, so the entire $32\pi\mu$ area increase must be due to rays entering the horizon through the caustic.
			
				Technically speaking, the line at which the null rays meet is not a caustic, but a \emph{crossover set} --- a set of points at which non-neighboring rays meet.  The crossover set traverses a line inside the small strip
				comprising the caustic horizon, and is terminated on both ends by caustic points --- points at which neighboring rays meet.  This is to the description given by Husa and Winicour \cite{Winicour} for asymmetric mergers.
				By extending the compact mass onto a stringlike object, we break the rotational symmetry of our problem, which in turn causes the horizon to be perturbed asymmetrically, forming caustics \emph{as well as} a crossover set.
				
			\subsection{Orthogonal Strings}
				Consider instead a string which falls into the black hole orthogonal to the event horizon.  In this case, returning to polar coordinates, we simply spread out the delta-function in (\ref{eq:caustic-th}),
				obtaining:
				\beq
					\left(\frac{d}{dv} - \frac{1}{4}\right) \frac{d}{dv} \theta = -\frac{\mu}{\theta} f(v)
				\eeq
				where $f(v)$ integrates to unity.  Now suppose that the string is very thin, so that the forcing term $f(v)$ is very small.  In this case, we may neglect $\ddot{\theta}(v)$ as small compared to $\dot{\theta}(v)$, yielding
				the following equation of motion:
				\beq
					\frac{d(\theta^2)}{dv} = 8\mu f(v)
				\eeq
				Rays will cross $\theta = 0$ and form caustics if and only if:
				\beq
					\theta_+ < \sqrt{8\mu}
				\eeq
				This is slightly larger than the caustic horizon found for a point particle.  It corresponds to a black hole area increase of:
				\beq
					\Delta A_c = 4(8\pi \mu) = 32\pi \mu
				\eeq
				Again, the caustic is responsible for 100\% of the black hole area increase, just as in the case of the parallel string.
				
				Other objects, including oblique strings and two-dimensional sheets, may be studied using the methods of this section.  Like parallel and orthogonal strings, such objects are expected to produce caustics which account for
				100\% of the black hole area increase.  The analysis, however, is less enlightening and is not presented here.
			
	\section{Conclusions} \label{sec:conclusions}
	
		In this paper, we used black-hole perturbation theory to model the event horizon of an extreme-mass-ratio merger. 
		While it is straightforward to obtain the  event-horizon deformation by numerically integrating the Zerilli and Regge-Wheeler equations --- and then geodesic equations, we have taken advantage of the extreme mass ratio and
		used an impulse approximation, which allowed us to find a universal geometry of the event horizon for all such mergers.

		While most of the large hole's horizon generators originate on the future horizon and remain on it as the black holes marge, a small subset originate on the \emph{past} horizon of the large black hole, wander away, and 
		are gravitationally lensed by the small black hole's field.  They subsequently enter the horizon at the \emph{caustic} and remain on the horizon after the merger.  The bundle of rays entering the horizon through the caustic
		is given by:
		\beq
			\theta < \sqrt{4\mu}
		\eeq
		where, without loss of generality, we assume the small black hole reaches the horizon at the $z$-axis.
		
		We find a caustic which is qualitatively similar to that seen in previous studies of head-on collisions.  Locally, the horizon is cone-shaped at the caustic, and the cone becomes increasingly pointed as the small black hole approaches the
		horizon.  We find that the cone angle can be related to the invariant length per unit time in a given time slicing, and the total invariant length is calculated to be $8\sqrt{\mu}$.  We find that the structure of the caustic is universal, depending only on the masses of the two merging black holes.  While the trajectory of the small black hole may affect the deformations in the bulk
		of the horizon, it does not materially affect the caustic.  On the other hand, cone-shaped caustics are not the only caustics that may be formed; if it were possible for a string-shaped object to fall into the black hole,
		it would produce two caustic points at the ends of the string, at which neighboring rays meet, and a crossover line at which non-neighboring rays meet.  Extending the point mass into a string breaks the isotropy of the
		problem and necessarily changes the shape of the caustic.
		
		Lastly, we found that the black hole area increase is half due to rays entering the horizon through the caustic, and half due to the expansion of rays in the vicinity of the caustic.  This property is also universal -- while
		different infall trajectories may result in differing degrees of shear in the bulk, they all produce identical results at and near the caustic, where all of the area increase happens.
		
		That said, there are a number of limitations to this study.  First, by restricting ourselves to first-order perturbations, our results are only valid when the gravitational field of the small hole is weak.  Thus, while 
		weakly lensed generators are correctly described using this model, stronly-lensed generators -- those with $\theta_+ \sim \mu$ -- are not.  Our perturbation theory cannot resolve the caustic at lengthscales comparable to
		the small hole's Schwarzschild radius, and it is plausible that at these lengthscales the conical structure of the caustic breaks down.  In order to go beyond the results of this paper, we would need to either consider higher
		orders in the perturbation expansion, or develop a qualitatively different technique for modeling the merger.
		
		Finally, while this paper is restricted to Schwarzschild black holes, most black holes in the real universe have spin.  Moreover, there does not exist a straightforward formalism for metric perturbations in the Kerr metric,
		so the approach taken by this paper is not easily transferrable to Kerr. Nevertheless, the dramatic simplifications and the universal geometrical features shown in this paper indicate that, for extreme mass ratios and for
		near-horizon geometry, it may not be necessary to start with the full metric-perturbation equations.  It is conceivable that a Rindler Approximation near the Kerr horizon would allow us to demonstrate the same caustic
		geometry (in a coordinate system co-rotating with the horizon) as in a plunge into Schwarzschild black holes. 

\begin{acknowledgments}

We thank Huan Yang and Kip S.\ Thorne for very useful discussions.  We thank Emanuele Berti for sharing with us his research notes on black-hole perturbation theory. Research of R.H. was supported by the David and Judith Goodstein Summer Undergraduate Research Fellowship  Endowment at Caltech.   Y.C.\ is supported by NSF grants PHY-0653653 and PHY-0601459, CAREER grant PHY-0956189, and the David and Barbara Groce start-up fund at Caltech. 	
\end{acknowledgments}
	\appendix

% A NOTE FOR FUTURE REFERENCE:
%
% This paper is the first in my series to make use of Zerilli's notation.  Before this paper, I made use of a hybrid of Berti's notation and my own.  Passing to Zerilli's notation required me to change some terms by
% a factor of i.  Here is the list of terms that were altered by passing to Zerilli's notation:
%
% A_{lm}^{(1)} -->  i A_{lm}^{(1)}
% B_{lm}^{(0)} -->  i B_{lm}^{(0)}
% Q_{lm}^{(0)} --> -i Q_{lm}^{(0)}
%
% All of the other terms are the same.  These are not very big changes, but one should make note of them when trying to compare the equation in this paper to any previous research documents.
% - R.H., 6/28/10
					
\section{Even-Parity Perturbations} \label{sec:even}

In the following two Appendices, we give the perturbed Einstein Equations and discuss how the metric perturbation is related to the Zerilli and Regge-Wheeler functions.  To do so, we use the formalism and results of Zerilli's paper~\cite{ZerilliPaper}.
		
		\subsection{Einstein Equations} \label{sec:eveneinstein}
			The perturbed Einstein Equations take the form:
			\beq
				\delta G_{\mu\nu} = 8\pi T_{\mu\nu}
			\eeq
			We express the energy-momentum tensor in terms of seven spherical harmonic components: $A_{lm}$, $A_{lm}^{(0)}$, $A_{lm}^{(1)}$, $B_{lm}$, $B_{lm}^{(0)}$, $F_{lm}$, and $G_{lm}$. 
			
%			\textcolor{red}{How difficult would it be if we follow Zerilli's convention instead of Berti's? Berti's note is unpublished, so it's not easy to get for the reader. It seems to me that there's only factors of 2, $i$, or $\sqrt{2}$ differences.}
%			Not very difficult.  But will take some time...			
			
			The tensor, expressed in matrix form, looks like the following:
			
			\begin{widetext}
			\beq
				T_{\mu\nu} =\frac{1}{\sqrt{2}} \left[\begin{array}{cccc} \sqrt{2 } A_{lm}^{(0)}Y^{lm} & i A_{lm}^{(1)}Y^{lm} & i r C_1 B_{lm}^{(0)}  Y^{lm}_{,\theta} & i r C_1 B_{lm}^{(0)}  Y^{lm}_{,\phi} \\ 
					{*} & \sqrt{2}  A_{lm} Y^{lm} & r  C_1 B_{lm} Y^{lm}_{,\theta} &  r C_1 B_{lm} Y^{lm}_{,\phi} \\
					{*} & * &  r^2( G_{lm} Y^{lm} + C_2  F_{lm} W^{lm}) & r^2 C_2  F_{lm} X^{lm} \\
					{*} & * & * &  r^2 \sin^2\theta ( G_{lm}  Y^{lm}  + C_2  F_{lm} W^{lm})
					\end{array}	\right]
			\eeq
%			\beq
%				T_{\mu\nu} = \left[\begin{array}{cccc} Y^{lm}A_{lm}^{(0)} & 2^{-1/2}Y^{lm}A_{lm}^{(1)} & * & 2^{-1/2}C_1 r Y^{lm}_{,\phi} B_{lm}^{(0)} \\ 
%					2^{-1/2}Y^{lm}A_{lm}^{(1)} & Y^{lm} A_{lm} & 2^{-1/2} C_1 r Y^{lm}_{,\theta}B_{lm} & {*} \\
%					{*} & C_1 r Y^{lm}_{,\theta}B_{lm} & 2^{-1/2} r^2(Y^{lm} G_{lm} + C_2 W^{lm} F_{lm}) & * \\
%					2^{-1/2}C_1 r Y^{lm}_{,\phi} B_{lm}^{(0)} & * & * & 2^{-1/2}r^2 \sin^2\theta (Y^{lm} G_{lm} + C_2 W^{lm} F_{lm}) \end{array}\right]
%			\eeq
Here we have defined 		
\beq
			C_1 = \frac{1}{\sqrt{2(\lambda+1)}},\quad C_2 = \frac{1}{2\sqrt{\lambda(\lambda+1)}}
\eeq
and 
		\begin{eqnarray}
			X^{lm} & = & 2\frac{\partial}{\partial\phi}\left(\frac{\partial}{\partial\theta} - \cot\theta\right) Y^{lm} \\
			W^{lm} & = & \left[\frac{\partial^2}{\partial\theta^2} - \cot\theta\frac{\partial}{\partial\theta} - \frac{1}{\sin^2\theta} \frac{\partial^2}{\partial\phi^2}\right] Y^{lm}
		\end{eqnarray}
The symbol * stands for terms obtainable through symmetry of $T_{\mu\nu}$.   Seven distinct Einstein
			Equations are found in all:
			\begin{eqnarray}
				-8\pi A_{lm}^{(0)} & = & \frac{(r-2)^2}{r^2}\frac{\partial^2 K}{\partial r^2} + \frac{(r-2)(3r-5)}{r^3} \frac{\partial K}{\partial r} - \frac{(r-2)^2}{r^3} \frac{\partial H_2}{\partial r} -
					\frac{r-2}{r^3}(H_2 - K) - \frac{(\lambda+1)(r-2)}{r^3}(H_2 + K) \nonumber \\
					\label{eq:b57} & & \\
				\label{eq:b58} -\frac{8\pi i}{\sqrt{2}} A_{lm}^{(1)} & = & \frac{\partial}{\partial t}\left[\frac{\partial K}{\partial r} + \frac{K - H_2}{r} - \frac{1}{r(r-2)} K\right] - \frac{\lambda+1}{r^2} H_1 \\
				-8\pi A_{lm} & = & \frac{r^2}{(r-2)^2} \frac{\partial^2 K}{\partial t^2} - \frac{r-1}{r(r-2)} \frac{\partial K}{\partial r} - \frac{2}{r-2}\frac{\partial H_1}{\partial t} + \frac{1}{r}\frac{\partial
					H_0}{\partial r} + \frac{1}{r(r-2)}(H_2 - K) + \frac{\lambda+1}{r(r-2)}(K - H_0) \nonumber \\
					\label{eq:b59} & & \\
				\label{eq:b60} 8\pi i \sqrt{2} r C_1 B_{lm}^{(0)} & = & \frac{r-2}{r} \frac{\partial H_1}{\partial r} + \frac{2}{r^2} H_1 - \frac{\partial(H_2 + K)}{\partial t} \\
				\label{eq:b61} 8\pi\sqrt{2} C_1(r-2) B_{lm} & = & -\frac{\partial H_1}{\partial t} + \frac{r-2}{r} \frac{\partial(H_0 - K)}{\partial r} + \frac{2}{r^2} H_0 + \frac{r-1}{r^2}(H_2 - H_0) \\
				\label{eq:b62} 8\pi\sqrt{2} C_2 r^2 F_{lm} & = & \frac{H_0 - H_2}{2} \\
			  8\pi\sqrt{2} G_{lm} & = & -\frac{r}{r-2} \frac{\partial^2 K}{\partial t^2} + \frac{r-2}{r} \frac{\partial^2 K}{\partial r^2} + \frac{2(r-1)}{r^2} \frac{\partial K}{\partial r}
					- \frac{r}{r-2}\frac{\partial^2 H_2}{\partial t^2} + 2\frac{\partial^2 H_1}{\partial r \partial t} - \frac{r-2}{r}\frac{\partial^2 H_0}{\partial r^2} \nonumber \\
				\label{eq:b63} & & + \frac{2(r-1)}{r(r-2)} \frac{\partial H_1}{\partial t} - \frac{r-1}{r^2}\frac{\partial H_2}{\partial r} - \frac{r+1}{r^2}\frac{\partial H_0}{\partial r} - \frac{\lambda+1}{r^2}(H_2 - H_0)
			\end{eqnarray}
		
		\subsection{The Zerilli Function} \label{sec:evenzer}
Instead of demonstrating how the Zerilli function is motivated, we simply write down its expression,\beq
				\label{eq:zer} Z = \frac{r^2}{\lambda r+3} K - \frac{r-2}{\lambda r+3} \int{H_1dt}
\eeq
and outline how to show that it satisfies a wave equation, and how to obtain all perturbation fields from it.  Throughout this section, we assume integrals in time range from $-\infty$ to $t$. We also assume all metric perturbation fields $(H_0,H_1,H_2,K)$ to vanish at $-\infty.$

In order to arrive at the wave equation, we will have to take $r$ derivatives and $t$ derivatives.  As it turns out, $r$ derivatives can be simplified when we repeatedly apply 
\beq
				\label{eq:b65} \frac{\partial K}{\partial r} = -\frac{r-3}{r(r-2)} K + \frac{1}{r} H_2 + \frac{\lambda+1}{r^2} \int{H_1dt} - 4\pi i \sqrt{2} \int{A_{lm}^{(1)}dt}\,,
\eeq
which can be obtained by integrating (\ref{eq:b58}) in time from $-\infty$ to $t$, and
\beq
				\label{eq:b66} 
				\frac{\partial}{\partial r}\int H_1 dt = \int{\frac{\partial H_1}{\partial r}dt} = \frac{2}{r(r-2)} \int{H_1dt} - \frac{r}{r-2}(H_2 + K) - \frac{8\pi i \sqrt{2} r^2 C_1}{r-2} B_{lm}^{(0)}\,,
\eeq
which can be obtained by integrating and simplifying (\ref{eq:b60}).
Incidentally, after substituting (\ref{eq:b65}) and (\ref{eq:b66}), $\partial Z/\partial r$ may also be expressed in terms of $K$ and $\int{H_1dt}$:
\beq
				\label{eq:dzer} \frac{\partial Z}{\partial r} = -\frac{r(\lambda r^2 - 3\lambda r - 3)}{(r-2)(\lambda r+3)^2} K + \frac{\lambda(\lambda+1)r^2 + 3\lambda r + 6}{r(\lambda r+3)^2} \int{H_1dt} 
					-\frac{4\sqrt{2}\pi i r^2}{\lambda r+3}\int{A_{lm}^{(1)}dt} - \frac{8\sqrt{2}\pi i r^2 C_1}{\lambda r+3}\int{B_{lm}^{(0)}dt}\,.
\eeq
Further calculation shows that $\partial^2 Z/\partial r_*^2$ is a combination of $K$, $\int H_1 dt$, $H_2$ and source terms.  

On the other hand, $\partial^2 Z/\partial t^2$ can be expressed also in terms of $K$, $\int H_1 dt$ and $H_2$, with the help of Eq.~\eqref{eq:b59}, plus
			\begin{eqnarray}
				\frac{\partial H_2}{\partial r} & = & -\frac{r-3}{r(r-2)} K + \frac{r-4}{r(r-2)} H_2 + \frac{r}{r-2}\frac{\partial H_1}{\partial t} + \frac{\lambda+1}{r^2} \int{H_1dt} \nonumber \\
				\label{eq:b68} & & -4\pi i\sqrt{2} \int{A_{lm}^{(1)}dt} - 16\pi\sqrt{2} C_2 \left[\frac{\partial(r^2 F_{lm})}{\partial r} - \frac{r(r-3)}{r-2} F_{lm}\right] + 8\pi\sqrt{2} C_1 r B_{lm}
			\end{eqnarray}
which can be obtained by substituting (\ref{eq:b62}) and (\ref{eq:b65}) into (\ref{eq:b61}).

Putting together $\partial ^2 Z/\partial r_*^2$ and $\partial^2 Z/\partial t^2$, we obtain
			\beq
				\label{eq:zereq} -\frac{\partial^2}{\partial t^2 }Z + \frac{\partial^2}{\partial r_*^2 }Z  - V_l^Z Z = S^Z_{lm}
			\eeq
			where
			\begin{eqnarray}
				V_l^Z & = & (1-2/r)\frac{2\lambda^2(\lambda+1)r^3 + 6\lambda^2 r^2 + 18\lambda r + 18}{r^3(\lambda r+3)^2} \\
				S^Z_{lm} & = & \frac{8 \pi(r-2)^2}{\lambda r+3} A_{lm} + \frac{8\pi(r-2)^2}{\sqrt{\lambda+1} (\lambda r+3)} B_{lm} - \frac{8\sqrt{2}\pi(r-2)}{\sqrt{\lambda(\lambda+1)}} F_{lm} - \frac{8\sqrt{2}\pi i (r-2)
						(r(\lambda+3) - 3)}	{r(\lambda r+3)^2} \int{{A_{lm}^{(1)}}dt} \nonumber \\
					\label{eq:zersrc} & + & \frac{8\pi i(r-2) (r^2 \lambda^2 + 3r(\lambda-2) + 12)}{\sqrt{\lambda+1} r(\lambda r+3)^2} \int{{B_{lm}^{(0)}}dt} - \frac{4\sqrt{2}\pi i(r-2)^2}{\lambda r+3} \int{{A_{lm}^{(1)}}_{,r}dt}
						\nonumber \\
					& - & \frac{8\pi i (r-2)^2}{\sqrt{\lambda+1}(\lambda r+3)} \int{{B_{lm}^{(0)}}_{,r}dt} 
			\end{eqnarray}
			Once the Zerilli function is known, the metric perturbations may be calculated accordingly.  Inverting  Equations (\ref{eq:zer}) and (\ref{eq:dzer}), we find:
			\begin{eqnarray}
				\label{eq:k} K & = & \frac{r^2\lambda(\lambda+1) + 3r\lambda + 6}{r^2(\lambda r+3)} Z +\frac{r-2}{r} \frac{\partial Z}{\partial r} + \frac{4\sqrt{2}\pi i r(r-2)}{\lambda r+3} \int{{A_{lm}^{(1)}}dt} + 
					\frac{8\sqrt{2}\pi i r(r-2) C_1}{3+r \lambda } \int{{B_{lm}^{(0)}}dt} \\
				\label{eq:h1} H_1 & = & \frac{r^2\lambda - 3r\lambda - 3}{(r-2)(\lambda r+3)} \frac{\partial Z}{\partial t}  + r \frac{\partial^2 Z}{\partial r \partial t} +
					\frac{4\sqrt{2}\pi i r^3}{\lambda r+3} A_{lm}^{(1)} + \frac{8\sqrt{2}\pi i C_1 r^3}{r\lambda+3} B_{lm}^{(0)}
			\end{eqnarray}
			Substituting these formulae for $H_1$ and $K$ into (\ref{eq:b65}), we find an expression for $H_2$ in terms of the Zerilli function:
			\begin{eqnarray}
				H_2 & = & -\frac{r^3\lambda^2(\lambda+1) + 3r^2\lambda^2 + 9r\lambda + 9}{r^2(\lambda r+3)^2} Z + \frac{r^2\lambda - r\lambda + 3}{r(\lambda r+3)}\frac{\partial Z}{\partial r} + (r-2)\frac{\partial^2 Z}{\partial r^2}
					\nonumber \\
					& & - \frac{8\sqrt{2}\pi i C_1 r(r^2\lambda(\lambda-1) + 6r(\lambda-1) + 15)}{(\lambda r+3)^2} \int{B_{lm}^{(0)}dt} + \frac{8\sqrt{2}\pi i C_1 r^2(r-2)}{\lambda r+3} \int{{B_{lm}^{(0)}}_{,r}dt} \nonumber \\
					\label{eq:h2} & & + \frac{4\sqrt{2}\pi i r(r^2\lambda + 6r - 6)}{(\lambda r+3)^2} \int{A_{lm}^{(1)}dt} + \frac{4\sqrt{2}\pi i r^2(r-2)}{\lambda r+3} \int{{A_{lm}^{(1)}}_{,r}dt}
			\end{eqnarray}
			\end{widetext}
			
		\subsection{Source Terms} \label{sec:evensource}
			The point mass of the small black hole traces out a trajectory $(T(\tau), R(\tau), \Theta(\tau), \Phi(\tau))$ in the Schwarzschild spacetime.  Because of the spherical symmetry of the problem, we may set 
			$\Theta(\tau) = \pi/2$ by an appopriate rotation.  The trajectory of the particle is characterized by two invariants: the specific  energy $E = (1-2/r)u^t$, and the specific angular momentum $L = r^2 u^{\phi}$.  The effective potential
			of the black hole is 
\begin{equation} 
U(r) = (1-2/r)(1+L^2/r^2)\,,
\end{equation}  
which determines the geodesic motion via
\begin{equation}
(dr/d\tau)^2 + U(r) =E^2
\end{equation}			
			
For a rest mass of $m_0$, the energy-momentum tensor of the particle is,
\beq
				T_{\mu\nu} = \frac{m_0}{r^2} \frac{dT}{d\tau} \frac{dx_\mu}{dt}\frac{dx_\nu}{dt} \delta(r-R(t))\delta^{(2)}(\Omega-\Omega(t))
\eeq
The even-parity content of this tensor corresponds to:
			\begin{eqnarray}
				\label{eq:srca} A_{lm} & = & \frac{m_0}{(r-2)^2}\sqrt{E^2-U[r]}\delta (t-T(r)) Y_{lm}^{*}(t) \\
				\label{eq:srca1} A_{lm}^{(1)} & = & \frac{\sqrt{2}m_0 E}{i r(r-2)}\delta (t-T(r)) Y_{lm}^{*}(t) \\
				\label{eq:srcb} B_{lm} & = & \frac{i \sqrt{2}C_1 m m_0 E L}{ r^2(r-2)} \delta(t-T(r)) Y_{lm}^{*}(t) \\
				\label{eq:srcb0} B_{lm}^{(0)} & = & \frac{\sqrt{2}C_1 m m_0 E L(r-2)}{r^4\sqrt{E^2-U[r]}} \delta(t-T(r)) Y_{lm}^{*}(t) \quad\quad\\
				\label{eq:srcf} F_{lm} & = & -\frac{m_0 C_2 L^2}{\sqrt{2} r^4\sqrt{E^2-U[r]}}\delta (t-T(r)) W_{lm}^{*}(t)
			\end{eqnarray}
			The spherical harmonic $Y_{lm}^{*}(t)$ refers to the value of $Y_{lm}^{*}$ at $(\Theta(t), \Phi(t))$, and is therefore a time-dependent quantity.  The same holds for $W_{lm}^{*}(t)$.

	\section{Odd-Parity Perturbations} \label{sec:odd}
		The odd-parity perturbations will give rise to three distinct Einstein Equations, one of which may be discarded as redundant after the Regge-Wheeler gauge is imposed.  There are two metric perturbations, $h_0$ and $h_1$,
		which may be written in terms of a Zerilli-type function (although its existence predates Zerilli's work\cite{ReggeWheeler}) $Q$, which in turn satisfies a sourced wave equation.
		
		\subsection{Einstein Equations} \label{sec:oddeinstein}
			As in the even-parity case, the perturbed Einstein Equations take the form $\delta G_{\mu\nu} = 8\pi T_{\mu\nu}$.  This time, however, the energy-momentum tensor takes the following form:
						\begin{widetext}	

			\beq
				T_{\mu\nu} = \frac{1}{\sqrt{2}}\left[\begin{array}{cccc} 0 & 0 & C_1r \csc(\theta)Y^{lm}_{,\phi}Q_{lm}^{(0)} & - C_1r \sin(\theta)Y^{lm}_{,\theta}Q_{lm}^{(0)} \\
					{*} & 0 & iC_1r \csc(\theta)Y^{lm}_{,\phi}Q_{lm} & -iC_1r \sin(\theta)Y^{lm}_{,\theta}Q_{lm} \\
					{*} & * & iC_2r^2 \csc(\theta)X^{lm} D_{lm} & -iC_2r^2 \sin(\theta)W^{lm} D_{lm} \\
					{*} & * & * & -iC_2r^2 \sin(\theta)X^{lm} D_{lm}
					\end{array}\right]
			\eeq
			As in the even-parity case, we have replaced some of the quantities with asterisks to simplify the expression.
			The two equations we choose to solve are the $\delta G_{r\phi}$ and $\delta G_{\theta\phi}$ equations, which take the following form:
			\begin{eqnarray}
				\label{eq:b21} 8\pi \frac{i C_1 r}{\sqrt{2}} Q_{lm} & = & -\frac{r}{2(r-2)}\left[\frac{\partial^2 h_1}{\partial t^2} - \frac{\partial^2 h_0}{\partial r \partial t} + \frac{2}{r}\frac{\partial h_0}{\partial t} + 
				\frac{2\lambda(r-2)}{r^3} h_1\right] \\
				\label{eq:b22} 8\pi \frac{i C_2 r^2}{\sqrt{2}} D_{lm} & = & -\frac{1}{2}\left[\frac{r-2}{r}\frac{\partial h_1}{\partial r} + \frac{2}{r^2} h_1 - \frac{r}{r-2}\frac{\partial h_0}{\partial t}\right]
			\end{eqnarray}
\end{widetext}		
		\subsection{The Regge-Wheeler Function} \label{sec:oddzer}
			The Regge-Wheeler function can be expressed in terms of $h_1$ as:
			\beq
				\label{eq:zerodd} Q = \frac{r-2}{r^2} h_1
			\eeq
			Like the Zerilli function, $Q$ obeys a simple wave equation with an analytic potential and source term.  The equation is given by:
			\beq
				\label{eq:zereqodd} \Box Q - V^Q_l Q = S^Q_{lm}
			\eeq
			where
			\begin{eqnarray}
				V_l^Q & = & \left(1 - \frac{2}{r}\right)\left[\frac{2(\lambda+1)}{r^2} - \frac{6}{r^3}\right] \\
				\label{eq:zersrcodd} S^Q_{lm} & = & -{8\sqrt{2}\pi i (r-2)C_2} \Bigg[\frac{\partial}{\partial r}\left(\frac{r-2}{r} D_{lm}\right)
				\nonumber\\&&\qquad\qquad\qquad\qquad - \frac{\sqrt{2\lambda}(r-2)}{r^2} Q_{lm}\Bigg]
			\end{eqnarray}
			By solving this wave equation, we find the Zerilli function, but what we are really after is the metric perturbation.  Substituting (\ref{eq:zerodd}) into (\ref{eq:b21}-\ref{eq:b22}), we may relate $h_1$ and $h_0$ to
			$Q$:
			\begin{eqnarray}
				\label{eq:h0odd} h_0 & = & \frac{r-2}{r} \int{Qdt} + (r-2)\int{\frac{\partial Q}{\partial r}dt} \nonumber \\&+& 8i\sqrt{2} \pi r(r-2) \int{D_{lm}dt} \\
				\label{eq:h1odd} h_1 & = & \frac{r^2}{r-2} Q
			\end{eqnarray}
			
		\subsection{Source Terms} \label{sec:oddsource}
			Following the methods of Section \ref{sec:evensource}, we can obtain the source terms for the odd perturbations.  It is only necessary to obtain $D_{lm}$ and $Q_{lm}$, which are provided below:
			\begin{eqnarray}
				\label{eq:srcd} D_{lm} & = & \frac{\sqrt{2}C_2 m m_0 L^2}{ r^4 \sqrt{E^2 - U(r)}} \delta(t-T(r)) {Y^{lm}_{,\theta}}^* \\
				\label{eq:srcq} Q_{lm} & = & -\frac{i\sqrt{2}C_1 m_0 L}{ r^2(r-2)} \delta(t-T(r)) {Y^{lm}_{,\theta}}^*
			\end{eqnarray}

	% -----------------------------------------------------------------------------------------------------------------------------------------------------------------------------------------------------------------------------
	% I need to change all of this to make it more consistent with the present notation.			??? I think I already did.
	\section{Behavior of Zerilli and Regge-Wheeler Functions} \label{sec:zerillibehavior}
			\begin{figure}[tb]
				\centering
					\includegraphics[width=1.00\columnwidth]{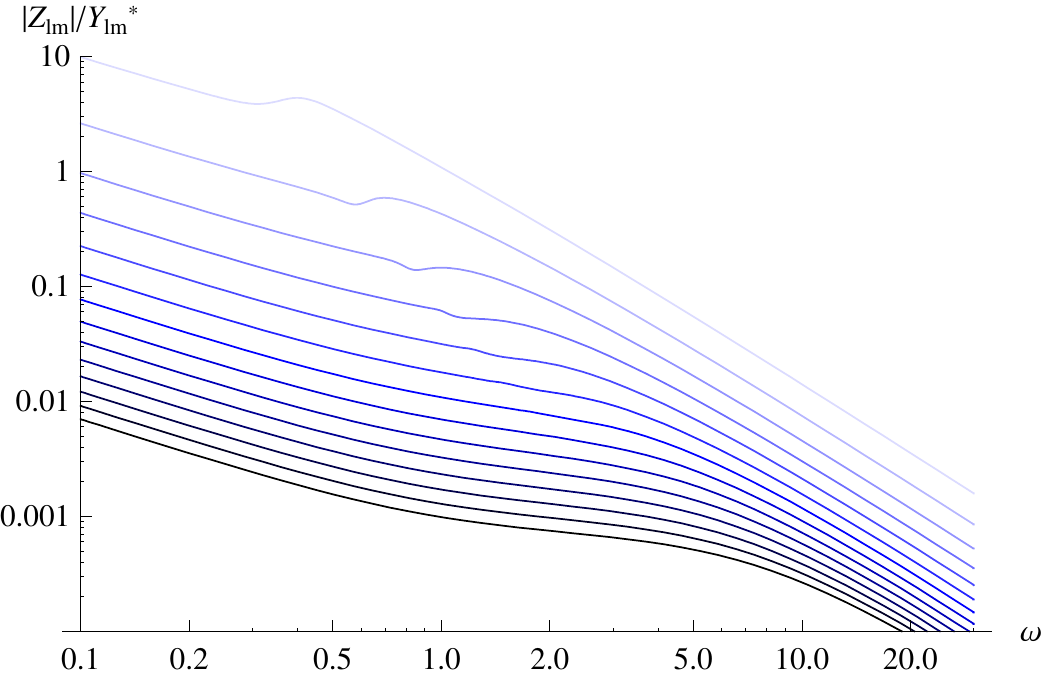}
				\caption{Plot of $|Z(\omega)|$ versus $\omega$.  Larger values of $l$ are denoted by darker lines.}
				\label{fig:ResearchThesis-1}
			\end{figure}
		The Zerilli and Regge-Wheeler functions are computed by Fourier transforming Equations (\ref{eq:zer}) and (\ref{eq:zerodd}) to the frequency domain and computing $Z(r_*, \omega)$ and $Q(r_*, \omega)$ using a standard ODE
		solver like Mathematica's \texttt{NDSolve[]}.  As per the standard practice, we impose an outgoing boundary condition at $r_* = \infty$ and an ingoing boundary condition at $r_* = -\infty$.  Since we are interested in the
		values of $Z$ and $Q$ at the horizon,
		we take $Z(r_*)$ (or $Q(r_*)$) with $r_*$ sufficiently negative (i.e. $r_* = -20$) that, for all practical purposes, it is on the horizon.  Once we have the horizon $Z(\omega)$ (or $Q(\omega)$), we Fourier transform it into
		the time domain, obtaining $Z(v)$ (or $Q(v)$).  Our Fourier transform algorithm takes into account the asymptotic behavior of $Z$ and $Q$ as $\omega \rightarrow 0$ and $\infty$.
		
		\subsection{Zerilli Function} \label{sec:zbehavior}
			In this section, we consider only the case of a radial infall.  The Zerilli function for a general infall will be qualitatively similar.
		
			Before calculating $Z(\omega)$, let us try to determine its general structure using intuition and some asymptotics.  Consider first, the case $\omega \rightarrow 0$.  This corresponds to the behavior of $Z$ on long
			timescales.  We know that for $v \rightarrow -\infty$, $Z \rightarrow 0$.  To get the limit as $v \rightarrow +\infty$, we relate $Z$ to the Zerilli-Moncrief function:~\cite{LoustoPrice}
			\beq
				Z_M = \frac{r}{\lambda + 1} \left[K + \frac{r-2}{\lambda r+3}\left(H_2 - r\frac{\partial K}{\partial r}\right)\right]
			\eeq
			\begin{figure}[tbp]
				\centering
					\includegraphics[width=1.00\columnwidth]{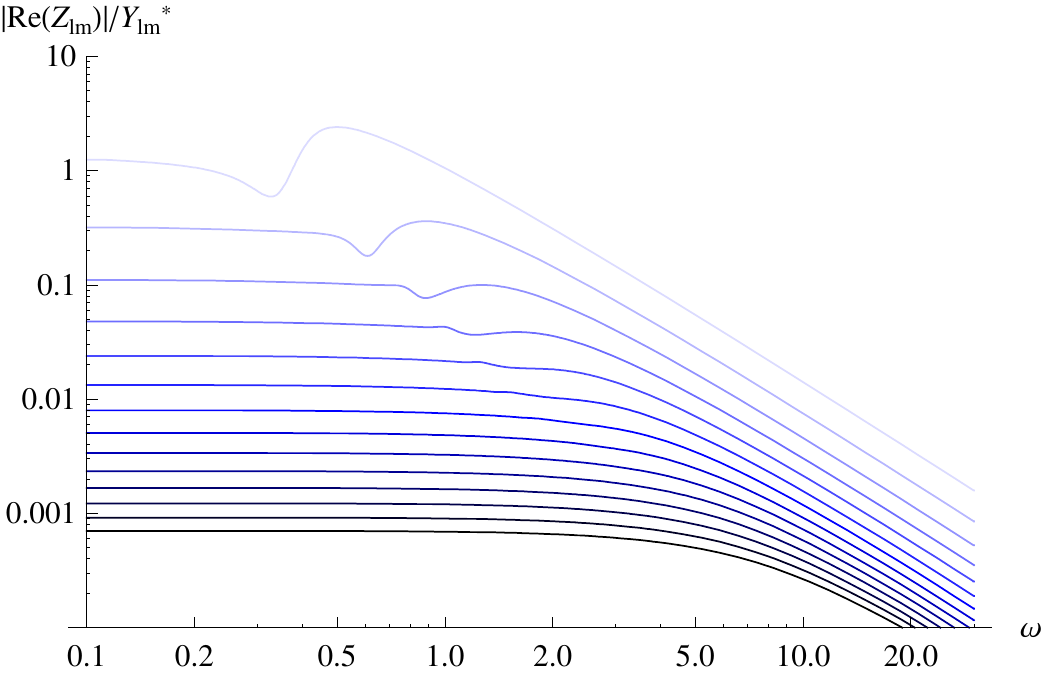}
				\caption{Plot of $|\mbox{Re}(Z(\omega))|$ versus $\omega$.  Larger values of $l$ are denoted by darker lines.}
				\label{fig:ResearchThesis-2}
			\end{figure}
			The Moncrief function is a linear combination of the metric perturbations and, by the No-Hair Theorem, must vanish as  $v \rightarrow +\infty$.  It is related to $Z$ by:
			\begin{eqnarray}
				Z_M & = & Z + \frac{4\sqrt{2}\pi i r^2(r-2)}{(\lambda + 1)(\lambda r + 3)}\int{A_{lm}^{(1)}dt} \nonumber \\
					& = & Z + \frac{8 \pi \mu r Y_{lm}^{*}}{(\lambda + 1)(\lambda r + 3)}
			\end{eqnarray}
			Near the horizon, therefore, the Zerilli function takes the following $v \rightarrow +\infty$ limit:
			\beq
				Z \rightarrow -\frac{8 \pi \mu r Y_{lm}^{*}}{(\lambda + 1)(2\lambda + 3)}
			\eeq
			Thus, on long timescales, $Z(v)$ is approximated by the Heaviside function $Z(v) \approx Z_\infty H(v)$.  Taking the Fourier transform, we find the lower asymptotic limit:
			\beq \label{eq:zer0}
				Z \rightarrow \frac{1}{\sqrt{2\pi} i\omega}\frac{8 \pi m_0 E r}{(\lambda + 1)(\lambda r + 3)}Y_{lm}^{*}\ \ \ \ \mbox{as}\ \omega \rightarrow 0
			\eeq
			Now consider the asymptotic behavior as $\omega \rightarrow \infty$.  This encodes the short-lengthscale behavior of the Zerilli function near the path of the particle.  On these lengthscales, we can assume that $Z$ is
			sourced solely by a delta-function component along the particle's trajectory; this gives rise to a \emph{discontinuity in the Zerilli Function's derivative}.  Near the horizon, the source term for $Z$ may be written in the
			following form:
			\beq
				S^Z = \frac{8\pi m}{2\lambda + 3} \int{\delta^{(2)}\left[x^a-x^a(\tau)\right]d\tau}\ Y_{lm}^{*}
			\eeq
			where $x^a(\tau)$ refers to the path of the particle.  Since the potential $V(r)$ vanishes near the horizon, the Zerilli Equation reduces to a free, sourced one-dimensional wave equation.  Solving this by the method of
			characteristics, we find the following jump in the derivative $\partial Z/\partial v$:
			\beq \label{eq:dzerjump}
				\Delta Z' \equiv \left.\frac{\partial Z_{lm}}{\partial v}\right|_+ - \left.\frac{\partial Z_{lm}}{\partial v}\right|_- = -\frac{8\pi m E}{2\lambda + 3} Y_{lm}^{*(0)}
			\eeq
			This gives rise to the following asymptotic behavior:
			\beq \label{eq:zerinf}
				Z \rightarrow \frac{1}{\sqrt{2\pi}\omega^2} \frac{8\pi m E}{2\lambda + 3} Y_{lm}^{*(0)}\ \ \ \ \mbox{as}\ \omega \rightarrow \infty
			\eeq
			The frequency-domain Zerilli function may be computed numerically by solving the equation:
			\beq \label{eq:zerfreq}
				\frac{\partial^2}{\partial r_*^2}Z + \omega^2 Z - V^Z Z = \tilde{S}^Z(\omega)
			\eeq
			We plot this function in Figures \ref{fig:ResearchThesis-1} and \ref{fig:ResearchThesis-2} and compare it to the asymptotic limits derived above.
			
			We wish to derive a scaling relation for the Zerilli function for large $l$.  The first step in doing so is to note that the jump in $Z'(v)$ scales as $l^{-2}Y_{lm}^{*(0)}$, whereas the long-timescale jump $Z_\infty$
			scales as $l^{-4}Y_{lm}^{*(0)}$.  For large $l$ values, it is the contribution of the former that dominates.  This contribution may be extracted from $Z(\omega)$ by taking the real part.  This is plotted in Figure
			\ref{fig:ResearchThesis-2}.
			
			The limiting value $Z_0 \equiv \mbox{Re}(Z(\omega=0))$ may be obtained by looking at the asymptotic behavior of (\ref{eq:zerfreq}).  Since $V \sim l^2$ and $\mbox{Re}(S) \sim l^{-2}Y_{lm}^{*(0)}$, it follows that for large
			$l$, $Z_0 \sim l^{-4}Y_{lm}^{*(0)}$.  This behavior is plotted in dashed lines in Figure \ref{fig:ResearchThesis-2}.
			
\begin{figure}[tbp]
				\centering
					\includegraphics[width=1.00\columnwidth]{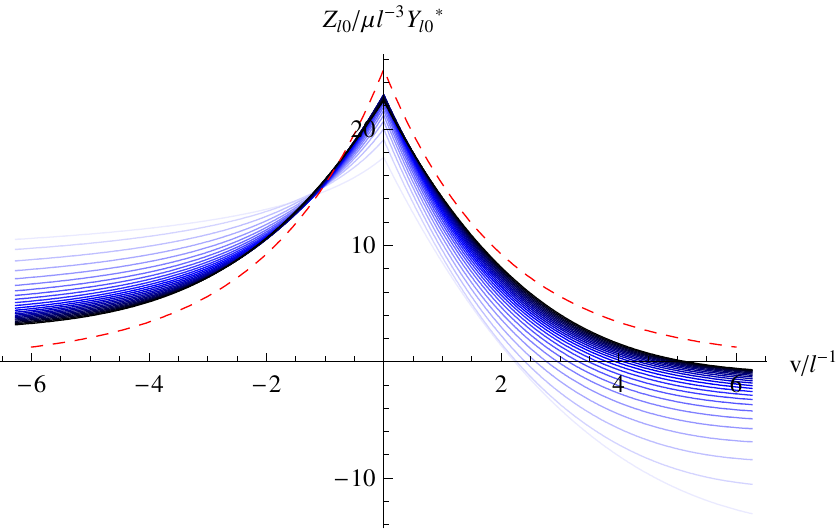}
				\caption{Plot of $Z(v)/l^{-3}Y_{lm}^{*(0)}$ versus $v/l^{-1}$.  Larger values of $l$ are denoted by darker lines.  The limit curve is shown in red.}
				\label{fig:ResearchThesis-3}
			\end{figure}

			The cutoff frequency for $|\mbox{Re}(Z(\omega))|$ may be obtained by equating the asymptotics, which scale as $l^{-4}Y_{lm}^{*(0)}$ and $\omega^{-2} l^{-2}Y_{lm}^{*(0)}$ respectively.  We find $\omega_c \sim l$, or
			correspondingly in the time domain, $t_c \sim l^{-1}$.  Taking the Fourier transform of $Z(\omega)$, the value of $Z(v=0)$ is proportional to the area under the curve of $Z(\omega)$.  This area, in turn, scales as 
			$Z_0 \omega_c \sim l^{-3} Y_{lm}^{*(0)}$,	so $Z(v=0) \sim l^{-3}Y_{lm}^{*(0)}$.  We thus obtain a \emph{scaling relation for $Z(v)$ for large $l$}:
			\beq
				\frac{Z_{lm}}{l^{-3}Y_{lm}^{*(0)}} = f(v/l^{-1})
			\eeq
			where $f(\xi)$ is an undefined function which is independent of $v$.  Plotting $Z_{lm}/l^{-3}Y_{lm}^{*(0)}$ against $v/l^{-1}$, we obtain the graph in Figure \ref{fig:ResearchThesis-3}.  Empirically, $Z(v)$ approaches an
			exponential:
			\beq \label{eq:zerasy}
				Z_{lm} \rightarrow 8\pi l^{-3} Y_{lm}^{*(0)} e^{-l|v|/2}
			\eeq
						
			What effect does this $Z(v)$ have on the horizon?  Recall that the metric perturbations act on the horizon through the forcing terms (\ref{eq:fr}-\ref{eq:fph}).  For even perturbations, all forcing terms are proportional
			to
			\beq
				f_{lm}^{(e)} = \frac{32\pi m_0 E Y_{lm}^{*(0)}}{2\lambda + 3} \delta(v) + 4\left(\frac{d}{dv} - \frac{1}{4}\right)\frac{dZ}{dv}
			\eeq
			Note that $f_{lm}^{(e)}$ contains two delta-functions -- one explicitly, and one due to $d^2 Z/dv^2$.  Noting the asymptotic form (\ref{eq:zerasy}) for $Z$, we can see that both of these delta-functions
			precisely cancel out!  The effect of the Zerilli function is to smooth out the delta-function in $f_{lm}^{(e)}$, resulting in a forcing term which is continuous and has a characteristic timescale $t_c \sim l^{-1}$.
			If we look on timescales much longer than $l^{-1}$, the forcing term may be approximated by a delta-function, as we showed in Section \ref{sec:delta-function}.
			
		\subsection{Regge-Wheeler Function} \label{sec:qbehavior}
			Like the Zerilli function, the Regge-Wheeler function $Q$ satisfies a wave equation with a very similar potential.  The source term (\ref{eq:zersrcodd}), however, is slightly different -- unlike the Zerilli
			function source, the $Q$ source does not contain a Heaviside-function component, and it does contain a $\delta'(t-T(r))$ component.  These differences will make the asymptotic behavior of $Q(\omega)$ slightly
			different from that of $Z$.
			
\begin{figure}[tbp]
				\centering
					\includegraphics[width=0.5\textwidth]{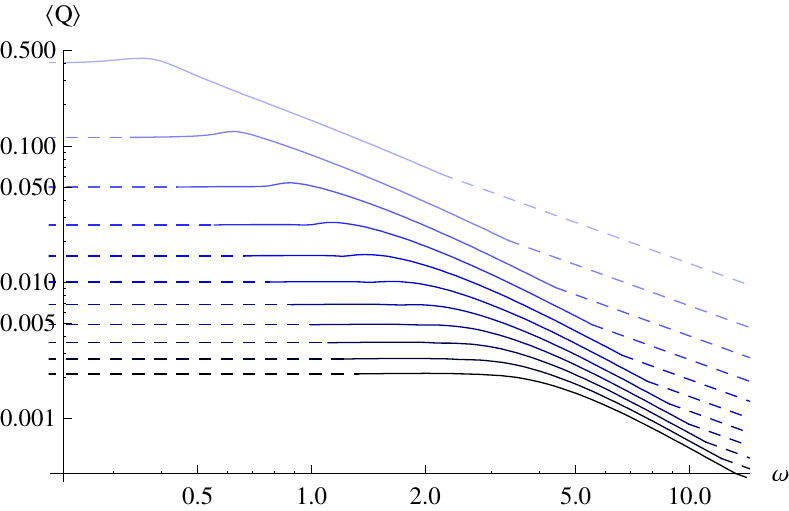}
				\caption{Plot of $\avg{Q(\omega)}$ versus $\omega$.  Larger values of $l$ are denoted by darker lines.}
				\label{fig:ResearchThesis-4}
			\end{figure}

			Like $Z(\omega)$, $Q$ is computed using a Mathematica ODE solver script.  We plot $Q$, rms averaged over $m$ values, in Figure \ref{fig:ResearchThesis-4}.
			
			We proceed to determine the general asymptotic behavior of $Q$.  It will turn out that $Q$'s contribution to the forcing terms is negligible near the caustic, so we do not take the time to derive exact formulas.
			For large $l$, the general behavior of $Q$ may be determined by an asymptotic analysis of the Zerilli Equation (\ref{eq:zereqodd}):
			\beq
				\frac{d^2}{dr_*^2} Q + (\omega^2 - V^Q) Q = S^Q
			\eeq
			In the limit $\omega \rightarrow 0$, the dominant term in $S^Q$ scales as $S^Q \sim l^{-2} {Y_{lm}^{*}}_{,\theta}$, while $V^Q \sim l^{2}$.  Since these terms dominate in the limit of large $l$, we find:
			\beq
				Q \sim \frac{S^Q}{V^Q} \sim l^{-4} {Y_{lm}^{*}}_{,\theta}\ \ \ \ \mbox{as}\ \omega \rightarrow 0
			\eeq
			Likewise, for large $\omega$, we must balance $\omega^2 Q$ on the left-hand side against $S^Q$ on the right-hand side.  The dominant part of $S^Q$ in this case scales as $l^{-4} m \omega {Y_{lm}^{*}}_{,\theta} \sim 
			l^{-3} \omega {Y_{lm}^{*}}_{,\theta}$.  Thus, the appropriate scaling for $Q$ is:
			\beq
				Q \sim \frac{l^{-3}}{\omega} {Y_{lm}^{*}}_{,\theta}\ \ \ \ \mbox{as}\ \omega \rightarrow \infty
			\eeq
			Like in the even case, the cutoff frequency for $Z_{odd}(\omega)$ scales as $\omega_c \sim l$ -- or conversely in the time domain, the odd-parity Zerilli function has a characteristic timescale $t_c \sim l^{-1}$.
			
			The odd-parity forcing terms from (\ref{eq:fr}-\ref{eq:fph}) are all proportional to:
			\beq
				f_{lm}^{(o)} = 8\left(\frac{d}{dv} - \frac{1}{4}\right)Q
			\eeq
			Like its even-parity equivalent, $f_{lm}^{(o)}$ has a characteristic timescale of $t_c \sim l^{-1}$.  If we look on timescales much longer than $l^{-1}$, the forcing term may be approximated by the delta-function
			derived in Section \ref{sec:delta-function}.  The odd-parity term turns out to be sub-dominant, and can accordingly be neglected.
			
		\section{Validity of Approximations} \label{sec:approx-validity}
			\subsection{Delta-Function Approximation}
				Recall from Appendix \ref{sec:zbehavior} that the Zerilli function scales as a function of $v/l^{-1}$, and therefore for large $l$, the forcing terms increasingly resemble delta-functions in time.  The
				impulse approximation, which treats these terms as delta-functions, results in a deflection of order $\Delta\theta \sim \mu^{1/2}$ for rays near the caustic horizon; as we show in this section, relative error in
				$\theta(v)$ incurred by imposing the impulse approximation is, at largest, of order $\mu^{1/4}$.  This is only an upper bound, however, and empirically the relative error in $\theta(v)$ appears to be of order $\mu^{1/2}$.
				Therefore, in the extreme mass-ratio case $\mu \ll 1$, deviations from the approximation may be safely ignored.
			
				\subsubsection{Proof by Asymptotics}
					First, we neglect the forcing term components $f_{lm}^{(e)}$ with $l \lesssim \mu^{1/4}$ and show that this introduces an error of order $\mu$.  Next, considering only components with $l \gtrsim \mu^{1/4}$, we show that
					the forcing terms may be written in terms of a smoothed-out delta function $f(v)$ with timescale $\tau \lesssim \mu^{1/4}$.  Deviations from the delta-function approximation give rise to an error at $\mu^{1/2}\tau
					\lesssim \mu^{3/4}$.
			
					Consider a particle falling in at the $z$-axis.  Recall from Section \ref{sec:delta-function} that for a given $l$, the even-parity forcing term component $f_{lm}^{(e)}$ takes the form:
					\begin{eqnarray}
						F_\theta & = & \frac{1}{16} f_{lm}^{(e)} Y_{lm,\theta} \nonumber \\
							& = & \frac{\pi m_0 E Y_{lm}^{*(0)}}{\lambda + 1} Y_{lm,\theta} \delta_l(v)
					\end{eqnarray}
					where $\delta_l$ is a delta-function spread out over a time interval of order $\tau_l \sim l^{-1}$ which integrates to unity.  First, we show that for generators in the vicinity of the caustic horizon 
					$\theta \sim \mu^{1/2}$, it is possible to ignore all terms with $l \leq \mu^{-1/4}$.  Re-expressing $Y_{l0}(\theta,\phi) \sim \sqrt{(2l+1)/4\pi} J_0(l \theta)$ for $\theta \ll 1$ (up to a relative error 
					$O(\theta) \sim O(\mu^{1/2})$), we can obtain the following expression for the forcing term:
					\begin{eqnarray}
						F_\theta & = & \sum_l{\frac{\mu (2l+1)}{2 l(l+1)} \delta_l(v) \frac{d}{d\theta} J_0(l\theta)} \nonumber \\
							& = & \sum_l{\frac{\mu (2l+1)}{2(l+1)} \delta_l(v) J'_0(l\theta)}
					\end{eqnarray}
					Now for $l \ll 1/\theta \sim \mu^{-1/2}$, $J'_0(l\theta) \sim l\theta/2 \sim m_0^{1/2} l$, giving a contribution to $F_\theta$ of $m_0^{3/2} l$.  Summing over all spherical harmonic terms with $l \leq m_0^{-1/4}$, we
					find a total contribution to $F_\theta$ of:
					\beq \label{eq:d3}
						F_{l \lesssim \mu^{-1/4}} \sim \mu^{3/2} \left(\mu^{-1/4}\right)^2 \sim \mu
					\eeq
					The total forcing term $F_\theta$ as derived in (\ref{eq:cfvec}), on the other hand, is of order $\mu/\theta \sim \mu^{1/2}$.  Therefore, the contribution \ref{eq:d3} of ``small-$l$'' ($l \lesssim \mu^{-1/4}$) terms
					is negligibly small.  We may therefore assume, to leading order in $\mu$, that only terms with $l \gtrsim \mu^{-1/4}$ contribute to the structure of the caustic.
				
					The forcing term, as a function of $v$, is thus concentrated within a small time interval of order $\tau \lesssim \mu^{1/4}$, and may be written to leading order in $m$ as $F_\theta = -(\mu/\theta) f(v)$, where $f(v)$
					integrates to unity.  Write $f(v)$ as a delta function plus a residual term: $\delta(v) + \epsilon \tilde{f}(v)$, where $\epsilon = 1$ is a placeholder.  Then $\theta(v)$ may be expanded in terms of $\epsilon$: 
					$\theta = \theta_0 + \epsilon \theta_1 + \ldots$.  Geodesic equation (\ref{eq:geo-th}), written as $\theta'' - \theta'/4 = -(\mu/\theta) (\delta(v) + \epsilon \tilde{f}(v))$ for the current forcing term, then breaks
					down into a series of equations:
					\begin{eqnarray}
						\theta_0'' - \frac{1}{4}\theta_0' & = & -\frac{\mu}{\theta_0} \delta(v) \\
						\theta_1'' - \frac{1}{4}\theta_1' & = & \frac{\mu}{\theta_0^2} \theta_1 \delta(v) + F(\theta_0) \tilde{f}(v)
					\end{eqnarray}
					We are not interested in the exact forms of $\theta_0$ and $\theta_1$.  Rather, we wish to determine how they vary to order of magnitude.  In the time interval $\tau \lesssim \mu^{1/4}$ over which the ray is lensed, we
					find that:
					\begin{eqnarray}
						\theta_0' & \sim & \frac{\mu}{\theta_0} \sim \mu^{1/2} \\
						\theta_0 - \theta_+ & \sim & \frac{\mu}{\theta_0}\tau \sim \mu^{1/2}\tau
					\end{eqnarray}
					Likewise, $\theta_1$ behaves as:
					\begin{eqnarray}
						\theta_1' & \sim & \int{\frac{\mu}{\theta_0^2} \theta_1 \delta(v) + \frac{\mu}{\theta_0} \tilde{f}(v)dv} \nonumber \\
						& \sim & \frac{\mu}{\theta_+^2} \theta_1\ \ \&\ \ \frac{\mu}{\theta_+^2} \int{(\theta_0 - \theta_+) \tilde{f}(v)dv} \nonumber \\
						& \sim & \theta_1' \tau\ \ \&\ \ \mu^{1/2} \tau
					\end{eqnarray}
					Thus, $\theta_1' \sim \mu^{1/2}\tau$.  This is a factor $\tau \lesssim \mu^{1/4}$ smaller than the deflection obtained using the delta-function approximation.  It follows that all deviations from the
					impulse appoximation are smaller than the delta-function result by a factor of $\tau \lesssim \mu^{1/4}$, and therefore the impulse approximation is valid for small $\mu$.
					
					The same does \emph{not} hold for the radial forcing term.  As we show in equation \ref{eq:cfr2}, the radial term $F^r$ scales as $O(m \log m)$, while the low-$l$ contributions to this term scale as $O(m)$.  However,
					the low-$l$ contributions are constant on the $O(m^{1/2})$ lengthscale associated with the caustic; this is best illustrated by noting that the spatial derivatives of $F^r$ are proportional to $F^\theta$ and $F^\phi$,
					which, as we shoed in this section, are well appoximated as impulses.  Thus, the impulse approximation for $F^r$ is valid up to a constant term $C^{(0)}(t)$, which depends on time but not position, near the caustic. 
					This constant term is a function of the low-$l$ perturbations and is $O(m)$.
			
				\subsubsection{Empirical Verification}
\begin{figure}
						\centering
							\includegraphics[width=1.00\columnwidth]{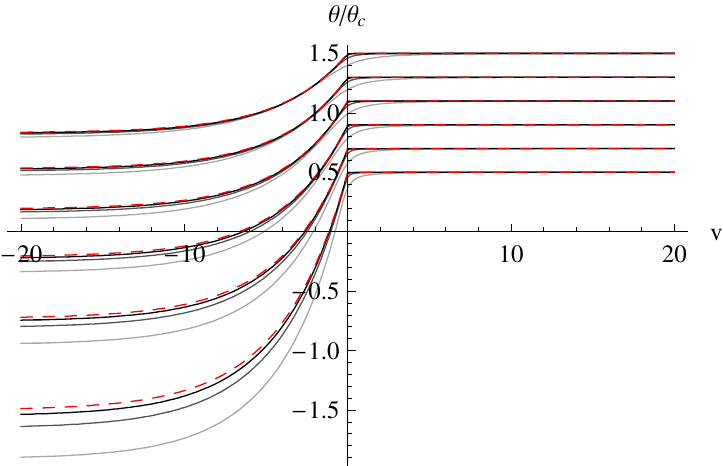}
						\caption{Plot of the horizon generators as lensed by Eq.\ \ref{eq:impulseshape2} for three masses: $\mu = 0.1$ (light gray), $0.01$ (dark gray), and $0.001$ (black), and the delta-function (red, dashed).}
						\label{fig:ResearchPaper2-Fig4}
					\end{figure}

					We can verify this empirically using the shape of the impulse, as derived in Eq.\ (\ref{eq:impulseshape}) of Section \ref{sec:impulseshape}:
					\beq \label{eq:impulseshape2}
						F^\theta(\theta, t) = -\frac{\mu}{4} \frac{\theta}{(\theta^2 + t^2/4)^{3/2}}
					\eeq
					To assess the validity of the impulse approximation, we propagate generators using two different forcing terms: the delta-function in (\ref{eq:dfth}) and the spread-out impulse in (\ref{eq:impulseshape2}).  This is 
					illustrated in Figure \ref{fig:ResearchPaper2-Fig4}, where we plot the paths of the generators for three masses: $\mu = 0.1$, $0.01$, and $0.001$.  As the particle's mass becomes smaller, the delta-function approximation
					becomes more accurate.
					
			  	For the two impulse shapes, we compute the displacement $\delta\theta = \left.\theta\right|^{+\infty}_{-\infty}$ and compare the two displacements.  Defining a \emph{relative error}
			  	\beq
				  	\epsilon \equiv \frac{|\delta\theta_1 - \delta\theta_2|}{\delta\theta_1}
			  	\eeq
			  	we find empirically that this error scales as:
			  	\beq
				  	\epsilon \sim \mu^{1/2} \left(\theta_c/\theta\right)
			  	\eeq
			  	Therefore, for rays near the caustic horizon, the relative error accrued by imposing the impulse approximation is $O(\mu^{1/2})$, well below the upper limit of $\mu^{1/4}$ derived in the previous section.
				
			\subsection{Neglecting Odd Perturbations}
				In general, the horizon generators are deflected by two types of forcing terms -- odd-parity and even-parity.  For rays near the caustic horizon, we show in this section that the ratio of odd-parity forcing to even-parity
				forcing is at most $O(\mu^{1/4})$.  Thus, for large mass ratios, the odd-parity terms may be neglected.
			
				The forcing terms may be written in terms of $f_{lm}^{(e)}$ and $f_{lm}^{(o)}$:
				\begin{eqnarray}
					F_\theta & \sim & f_{lm}^{(e)} Y^{lm}_{,\theta}\ \ \&\ \ f_{lm}^{(o)} Y^{lm}_{,\phi} \\
					F_\phi & \sim & f_{lm}^{(e)} Y^{lm}_{,\phi}\ \ \&\ \ f_{lm}^{(o)} Y^{lm}_{,\theta}
				\end{eqnarray}
				Accordingly, for a given $l$, the relative importance of odd and even perturbations is given by:
				\beq
					\frac{\mbox{Odd}_{lm}}{\mbox{Even}_{lm}} \sim \frac{f_{lm}^{(o)}}{f_{lm}^{(e)}} \sim \frac{m_0 l^{-4} Y_{lm,\theta}^*}{m_0 l^{-2} Y_{lm}^*} \sim l^{-1}
				\eeq
				Therefore, $f_{lm}^{(o)} \lesssim f_{lm}^{(e)}$ for all $l$.  Consider a null generator in the vicinity of the caustic horizon -- $\theta \sim \mu^{1/2}$.  As we showed in the previous subsection, the even-parity
				solution is dominated by terms with $l \gtrsim \mu^{-1/4}$.  The even-parity contribution for $l \lesssim \mu^{-1/4}$ is of order $\mu^{1/2}$ smaller than the total contribution, and therefore can be neglected.  Since
				$f_{lm}^{(o)} \lesssim f_{lm}^{(e)}$, it follows that the \emph{odd-parity} contribution of terms with $l \lesssim \mu^{-1/4}$ can be likewise neglected.  Odd terms for which $l \gtrsim \mu^{-1/4}$
				can likewise be neglected to leading order in $\mu$ because $f_{lm}^{(o)} / f_{lm}^{(e)} \sim l^{-1} \lesssim \mu^{1/4}$.
				
				It follows that the odd perturbations as a whole may be neglected.  The total odd-parity contribution to the forcing term is insignificant compared to the even-parity contribution:
				\beq
					\frac{\mbox{Odd}}{\mbox{Even}} \lesssim \mu^{1/4} \ll 1
				\eeq


\begin{thebibliography}{99}
			\bibitem{GW}
				K.S.\ Thorne, in {\it 300 Years of Gravitation}, Ed.\ S.W.\ Hawking and W.\ Israel, Cambridge University Press, 1987.
			\bibitem{PN}
				L.\ Blanchet, {\it Gravitational Radiation from Post-Newtonian Sources and Inspiralling Compact Binaries}, Living Rev. Relativity {\bf 5}, (2002), \arxiv{gr-qc/0202016}.
			\bibitem{BHP}
				A.\ Nagar and L.\ Rezzolla, Class.\ Quantum Grav.\ {\bf 22}, R167 (2005), \arxiv{gr-qc/0502064}; M.\ Sasaki and H.\ Tagoshi, {\it Analytic Black Hole Perturbation Approach to Gravitational Radiation},  
				Living Rev.\ Relativity {\bf 6}, (2003), \arxiv{gr-qc/0306120}.
			\bibitem{Pretorius}
				F.\ Pretorius, in {\it Relativistic Objects in Compact Binaries: From Birth to Coalescence}, Ed.\ M.\ Colpi et al., Springer, 2007.
			\bibitem{HF}
				E.E.\ Flanagan and S.A.\ Hughes, Phys.\ Rev.\ D {\bf 57}, 4535 (1998), \arxiv{gr-qc/9701039}
			
			\bibitem{Lovelace:2009}  G.\ Lovelace et al.\ (2009), \arxiv{0907.0869}.
			\bibitem{Keppel:2009} D.\ Keppel, D.A.\ Nichols, Y.\ Chen, and K.S.\ Thorne, \prd {\bf 80} 124015 (2009), \arxiv{0902.4077}.
			\bibitem{Rezzolla:2010} L.\ Rezzolla, R.P.\ Macedo and J.L.\ Jaramillo, \prl {\bf 104}, 221101 (2010), \arxiv{1003.0873}.
			\bibitem{Nichols:2010} D.\ Nichols and Y.\ Chen, (2010), \arxiv{1007.0204}.
			
				
			\bibitem{HawkingPaper}
				S. Hawking, \emph{Black Holes in General Relativity}, Comm. Math. Phys., \textbf{25}, 152 (1972), \doi{10.1007/BF01877517}
			\bibitem{Winicour}
				S.\ Husa and J.\ Winicour, Phys.\ Rev.\ D {\bf 60}, 084019 (1999), \arxiv{gr-qc/9905039}.
			\bibitem{HeadOn} R.\ Matzner et. al., Science \textbf{270} (1995), \doi{10.1126/science.270.5238.941}
			\bibitem{Libson:1996} J.\ Libson, J.\ Masso, E.\ Seidel, W.M.\ Suen, P.\ Walker, \prd {\bf 53} 4335 (1996), \arxiv{gr-qc/9412068}.
			\bibitem{Masso:1999} J.\ Masso, E.\ Seidel, W.M.\ Suen, P.\ Walker, \prd {\bf 59} 064015 (1999), \arxiv{gr-qc/9804059}. 
			\bibitem{CPS}
	  		Cohen, Pfeiffer and Scheel, Class. Quant. Grav. \textbf{26}:035005 (2009), \arxiv{0809.2628}.
			\bibitem{PenroseBook}
	  		Penrose, \emph{Techniques of Differential Topology in Relativity}, Regionial Conference Series in Applied Mathematics (1972).
	  	\bibitem{SiinoPaper}
	  		M.\ Siino, Phys. Rev. \textbf{D58} (1998), \arxiv{gr-qc/9701003}.
			\bibitem{Hartle:1974} J.B.\ Hartle, \prd {\bf 9} 2749 (1974), \doi{10.1103/PhysRevD.9.2749}
\bibitem{Poisson:2010} E.\ Poisson and I.\ Vlasov, \prd {\bf 81} 024029 (2010), \doi{10.1103/PhysRevD.81.024029}
\bibitem{Comeau:2009} S.\ Comeau and E.\ Poisson, \prd {\bf 80} 087501 (2009), \arxiv{0908.4518}.
\bibitem{Poisson:2009} E.\ Poisson, \prd {\bf 80} 064029 (2009).
\bibitem{Taylor:2008} S.\ Taylor and E.\ Poisson \prd {\bf 78} 084016 (2008). 
\bibitem{Preston:2006} B.\ Preston and E.\ Poisson \prd {\bf 74} 064009 (2006), \arxiv{gr-qc/0606093}.
\bibitem{Preston:2006b} B.\ Preston and E.\ Poisson \prd {\bf 74} 064010 (2006), \arxiv{gr-qc/0606094}.
\bibitem{Fang:2005} H.\ Fang and G.\ Lovelace \prd {\bf 72} 124016 (2005), \arxiv{gr-qc/0505156}. 
\bibitem{Poisson:2004} E.\ Poisson \prd {\bf 70} 084044 (2004), \arxiv{gr-qc/0407050}
\bibitem{Poisson:1995} E.\ Poisson and M.\ Sasaki \prd {\bf 51}, 5753 (1995), \arxiv{gr-qc/9412027}.
%	  	\bibitem{MatznerPaper}
%	  		Matzner et al., Science \textbf{270} (1995).
\bibitem{BornApproximation}
See, e.g., C.\ Cohen-Tannoudji, B.\ Diu and F.\ Lalo\"e, Quantum Mechanics, Sec.\ VIII\,B.
	  	\bibitem{ReggeWheeler}
	  		T.\ Regge and J. Wheeler, Phys. Rev. \textbf{108} (1957), \doi{10.1103/PhysRev.108.1063}
	  	\bibitem{ZerilliLetter}
	  		F.\ Zerilli, Phys. Rev. Lett. \textbf{24} (1970), \doi{10.1103/PhysRevLett.24.737}
	  	\bibitem{ZerilliPaper}
	  		F.\ Zerilli, Phys. Rev. \textbf{D2} (1970), \doi{10.1103/PhysRevD.2.2141}
	  	\bibitem{LoustoPrice}
	  		C.O.\ Lousto and R.H.\ Price, Phys. Rev. D \textbf{55}, 2124 (1997), \arxiv{gr-qc/9609012v1}.
	  	\bibitem{Eddington}
	  		F.W.\ Dyson, A.S.\ Eddington and C.\ Davidson, Phil.\ Trans.\ Roy.\ Soc.\ A {\bf 220}, 291-333 (1920), \doi{10.1098/rsta.1920.0009}
	  		
	\end{thebibliography}
\end{document}